\documentclass[twocolumn,twocolappendix]{aastex631}

\usepackage{color}
\definecolor{red}{rgb}{0.7,0.1,0.1}
\definecolor{blue}{rgb}{0.2,0.2,0.8}
\definecolor{green}{rgb}{0.1,0.6,0.1}

\long\def\Ignore#1{\relax}

\begin{document}

\title{Star Formation and Magnetic Field Amplification due to Galactic Spirals}

\author{Hector Robinson}
\affiliation{Department of Physics and Astronomy, McMaster University, 1280 Main Street West, Hamilton, Ontario, L8S 4M1 Canada}
\correspondingauthor{Hector Robinson}
\email{robinh4@mcmaster.ca}
\author{James Wadsley}
\affiliation{Department of Physics and Astronomy, McMaster University, 1280 Main Street West, Hamilton, Ontario, L8S 4M1 Canada}
\author{J. A. Sellwood}
\affiliation{Steward Observatory, University of Arizona,
933 Cherry Avenue,
Tucson, AZ 85722, USA}
\author{Ralph E. Pudritz}
\affiliation{Department of Physics and Astronomy, McMaster University, 1280 Main Street West, Hamilton, Ontario, L8S 4M1 Canada}



\begin{abstract}

We use global MHD galaxy simulations to investigate the effects of spiral arms on the evolution of magnetic fields and star formation within a self-regulated interstellar medium (ISM). The same galaxy is simulated twice: once with self-consistent stellar spiral arms and once more with the stellar spirals suppressed via a novel numerical approach, using the \textsc{Ramses} AMR code. Spiral arms continually promote star formation, with 2.6 times higher rates in the spiral galaxy.  The higher rate is due to high gas columns gathered along the spiral arms, rather than increasing the star formation efficiency at a given gas column. In both cases, the magnetic field is initially amplified via a small-scale dynamo driven by turbulence due to supernova feedback.  Only the spiral galaxy exhibits late-time, consistent field growth due to a large-scale dynamo (e-folding time $\sim600$ Myr).  This results in volume-averaged field strengths of $\sim 1$ $\mu$G after 1 Gyr of evolution. The mean-fields tend to align themselves with the spiral arms and are coherent up to 10 kpc scales. We demonstrate a novel large-scale dynamo mechanism, whereby spiral-driven radial flows enable the mean-field amplification. 
\end{abstract}



\section{Introduction} \label{sec:intro}

Spirals drive many aspects of evolution within disk galaxies.  They cause radial migration of stars and gas \citep{sellwood2002, Daniel2018}, flatten rotation curves \citep{lovelace1978, Berrier2015}, promote star formation \citep{elmegreen1986,Kim2020}, and drive turbulence throughout the ISM \citep{Kim2006}. Amplification of a galaxy's magnetic field is also closely related to its star formation, turbulence, and rotation. Thus, it seems likely that spiral arms assist the growth of magnetic fields through these effects.

Observations of magnetic fields in spiral galaxies have begun to constrain the field strengths and orientations. Synchrotron emission suggests fields $\sim$10 $\mu$G, that may be stronger in high density regions \citep{fletcher2011,basu,Beck}. Polarization measurements have shown that the fields tend to be organized into large-scale spirals, regardless of the presence of optical spirals \citep{chyzy2008,Beck2019,lopez-rodriguez}.  Other authors find that magnetic fields are generally aligned with, or slightly inclined to, gas structures down to 100 pc scales \citep{goldsmith2008, Frick2016, planck1, su2018, Beck2020, Zhao2024}.

In order to have reached detectable levels, magnetic fields in galaxies need to have been amplified through some dynamo process \citep{Geach2023, ledos2024}. Dynamos are typically classified into two main categories: small-scale and large-scale. Small-scale, or turbulent, dynamos amplify field strengths exponentially through the random stretching, twisting and folding of field lines in a turbulent medium  and tend to allow fast growth on smaller scales only. This process operates on short ($\lesssim10$ Myr) timescales but leaves the field disordered as viewed on larger scales.  Many simulations have demonstrated that a turbulent dynamo can generate field strengths approaching those observed in nature \citep{fed_ssdynamo, Bhat2016, Rieder1, pakmor, martin2022}.


Large-scale or mean-field dynamos are required to create fields that are coherent over galactic scales.  In principle, this could be due to driving a turbulent dynamo on the circumgalactic medium (galactic halo) scale \citep{Rieder2,pudritz1989}. The classical large scale dynamo is assumed to work through an interplay between galactic rotation and the buoyancy of the fields \citep{Parker1955}, a process that is predicted to operate on Gyr timescales. Mean-field dynamos for galaxies are commonly of the $\alpha - \Omega$ type \citep{brandenburg2005}.  The  $\alpha - \Omega$ dynamo is predicted to produce quadrupolar fields in the galactic halo, magnetic spirals, and helical turbulence \citep{stix1975, henriksen2017, shukurov2019}, some of which were observed in simulations by \citet{ntormousi2020}. 

Small-scale dynamos generate small-scale (also called {\it turbulent}) fields, while large-scale dynamos amplify the mean field.  Since galaxies contain both turbulent and mean fields we infer that the fields are amplified by some combination of small and large-scale dynamos.

Regardless of the scale, all proposed dynamos rely on the presence of turbulence within the system. Supernova feedback in star-forming disks is a robust source of turbulence that stirs the ISM and has been shown to be directly coupled to the amplification rate of the fields \citep{Rieder1, butsky2017}.  Supernova-driven turbulence tends to be isotropic and distributed throughout the ISM, making it an effective driver of the small-scale dynamo. In addition to stellar feedback, several instabilities may generate turbulence important to dynamo action, such as the magnetorotational instability (MRI) \citep{balbus1991,sellwood1999,korpi2010,gressel2013}, gravitational instability including spiral arms themselves\citep{toomre1964,krumholz2018,pfrommer2022}, and cosmic ray instabilities \citep{bell2004,riquelme2009}.

\citet{sellwood2002} showed how transient spiral patterns in a galaxy disk can cause efficient radial mixing of the stars within the disk.  As a spiral instability grows, the non-axisymmetric potential well first traps stars near corotation onto orbits that cross corotation.  Trapped stars may cross the resonance more than once as they alternately gain and lose angular momentum.  Later, as the disturbance decays, trapped stars may be released at new radii with substantially changed angular momenta.   A succession of transient spiral patterns having differing corotation radii can cause stars to random walk in several steps over large radial distances from their birth radii, a process that was also studied by \citet{Roskar2008} and by \citet{Daniel2018} and has been shown by \citet{Frankel2020}, for example, to have been an important driver of evolution in the Milky Way disk over the past 6 Gyr.

While radial migration due to spirals was first understood as a stellar dynamical phenomenon, \citet{sellwood2002} argued that clouds of gas should exhibit similar behavior, i.e. they become trapped by the gravitational potential of a growing spiral, cross corotation, and then are released in a similar manner to the stars.  
Magnetized gas will drag its entrained magnetic field, which will stretch the field in the radial direction at first, after which it will be sheared by differential rotation, while parcels of magnetized gas from different radii become mixed, thereby stirring the medium on much grander scales than is possible by supernova-driven turbulence, for example.  A more general version of this picture would allow for field to be dragged by spiral-linked gas flows without necessarily requiring compact clouds.

The extent of radial flows caused by spirals naturally depends on their peak amplitude, and also the multiplicity of the spiral pattern.  Several studies of magnetized spiral galaxies have employed submaximum disks \citep[e.g.][]{wibking2023,Robinson2024}, causing them to prefer multi-arm spirals \citep{Sellwood2022} that drive radial flows over shorter distances.  Simulations by \citet{Khoperskov2018} and by \citet{Kim2006}  directly investigated the relationship between spiral arms and magnetic field amplification and found that galaxies produce stronger mean-fields within spiral arms. \citet{ntormousi2020} found evidence for a mean-field dynamo operating in simulations of a forming spiral galaxy, but did not investigate the role of the spiral arms directly.  Altogether this motivates simulations that can clearly demonstrate the role spiral arms play in galactic magnetic field evolution.


Another effect commonly associated with spiral patterns is the promotion of star formation.  A key question is whether they  {\it trigger} star formation at increased efficiency or {\it gather} gas that forms stars at typical rates \citep{elmegreen1986,Sellwood2022}. Recent observational results have begun to favor the `gather picture, where spiral arms bring together gas that is already going to undergo star formation, but affect the overall efficiency by negligible amounts \citep{Sun2024, Querejeta2024}. The stratified sheared patch simulations of \citet{Kim2020} also favor the gather picture.

In this work, we compare simulations of an isolated galaxy with and without spiral arms, but with setups and physics that are otherwise identical. The spiral arms form as natural instabilities in the first simulation (i.e.\ they are not imposed as a fixed potential), which properly captures their transient nature.  Comparison of the two simulations then allows  us to study the role of spirals in both magnetic field amplification, and in the triggering/gathering of star formation. 

The remainder of this paper is organized as follows: In Section \ref{sec:methods} we describe our numerical methods and simulation setup, in Section \ref{sec:results} we present the results of the simulations, in Section \ref{sec:discussion} we discuss the implications, and in Section \ref{sec:conclusions} we make some broad conclusions.

\section{Methods} \label{sec:methods}
We conduct MHD simulations of two initially identical, isolated galaxies, but allow spiral arms to develop in one and inhibit their formation in the other.  Our methods mostly follow \cite{Robinson2024}. We use the \textsc{Ramses} adaptive mesh refinement (AMR) code \cite{RAMSES} that solves the ideal MHD equations using a Harten-Lax-van Leer discontinuities (HLLD) approximate Riemann solver.  We use the constrained transport method \citep{constrained_transport} to maintain the solenoidal constraint ($\nabla \cdot \mathbf{B}=0$), and a particle-mesh technique \citep{pm} to determine the gravitational attraction of stars, dark matter and gas.  Cooling and heating processes in the gas are accounted for through the \textsc{Grackle} chemistry and cooling library \citep{grackle}, which utilizes metal cooling rates from the photo-ionization code \textsc{Cloudy} \citep{Cloudy}.  We include a photoelectric heating rate of $\zeta = 4 \times 10^{-26}$ erg cm$^{-3}$ s$^{-1}$.

Star particles are formed stochastically with a Schmidt-Law of the form:
\begin{equation}
        \frac{d \rho_*}{dt} = \frac{\epsilon_{\rm ff}\rho}{ \, t_{\rm ff}} \quad \text{if} \quad \rho > \rho_{\rm crit}
\end{equation}
where $\rho_*$ is the stellar density, $\rho_{\rm crit}$ is a threshold density which corresponds to a number density of 100 cm$^{-3}$, and $\epsilon_{{\rm ff}}$ is the star formation efficiency per free-fall time. To reduce the magnitude of an initial starburst, we initially set $\epsilon_{\rm ff}$ to 0.01, and gradually increase it to 0.1 over the first 200 Myr of evolution to suppress any initial starburst.

In order to mimic supernova feedback, we assume each newly formed star particle releases $10^{51}$ erg per 91 M$_{\odot}$, after a 5 Myr delay.  We prevent the released energy from cooling at first using the delayed cooling model of \cite{delayed_cooling}, which allows unresolved superbubbles to grow properly, but it is then converted into regular thermal energy with an e-folding timescale of 5 Myr.  The gas disk is initialized inside of a domain 600 kpc on a side that has a base grid of 64$^3$ cells, that is allowed to refine an additional 10 levels which gives a spatial resolution of 9.15 pc at the highest level.

\subsection{Galaxy models with and without spirals}
We created a self-consistent, equilibrium dark matter halo and the initial stellar disk as described in the Appendix and use it for both simulations.  

We allow spirals to develop in one model, by evolving the stellar disk and DM halo for 400 Myr without gas.  Once spiral arms have developed, we subtract 10\% from the masses of the disk star particles and convert that mass to gas in an exponential disk with a scale-height and scale-radius of 0.34 kpc and 3.4 kpc respectively.  In this way, we begin the full MHD simulation with a stellar disk that has already developed spiral arms.  The initial magnetic field embedded within the gas disk is purely toroidal with a field strength 
\begin{equation}
    B = 0.85 \left(\frac{\rho}{\rho_0}\right)^{2/3}\ \mu{\rm G},\label{Binit} 
\end{equation}
where $\rho_0$ = 1.15 $\times$ 10$^{-23}$ g cm$^{-3}$ is the initial density at the galaxy center.  These choices result in a magnetic field strength of 0.1 $\mu$G in a typical ISM number density of 0.25 cm$^{-3}$. 

To inhibit stellar spirals from forming in the second model we measured the initial axisymmetric values of the gravitational accelerations of the disk star particles as a function of radius, and modified \textsc{Ramses} to continuously use those values rather than those calculated by the gravity solver.  This forces the stellar disk to remain axisymmetric and prevents any stellar spiral arms from forming. The gas and DM halo continue to respond to their own self-gravity and that of the stellar disk particles. 

From this point on we shall refer to the two models as the spiral galaxy, and the no-spiral galaxy.

\begin{figure}[ht!]
\includegraphics[width=\columnwidth]{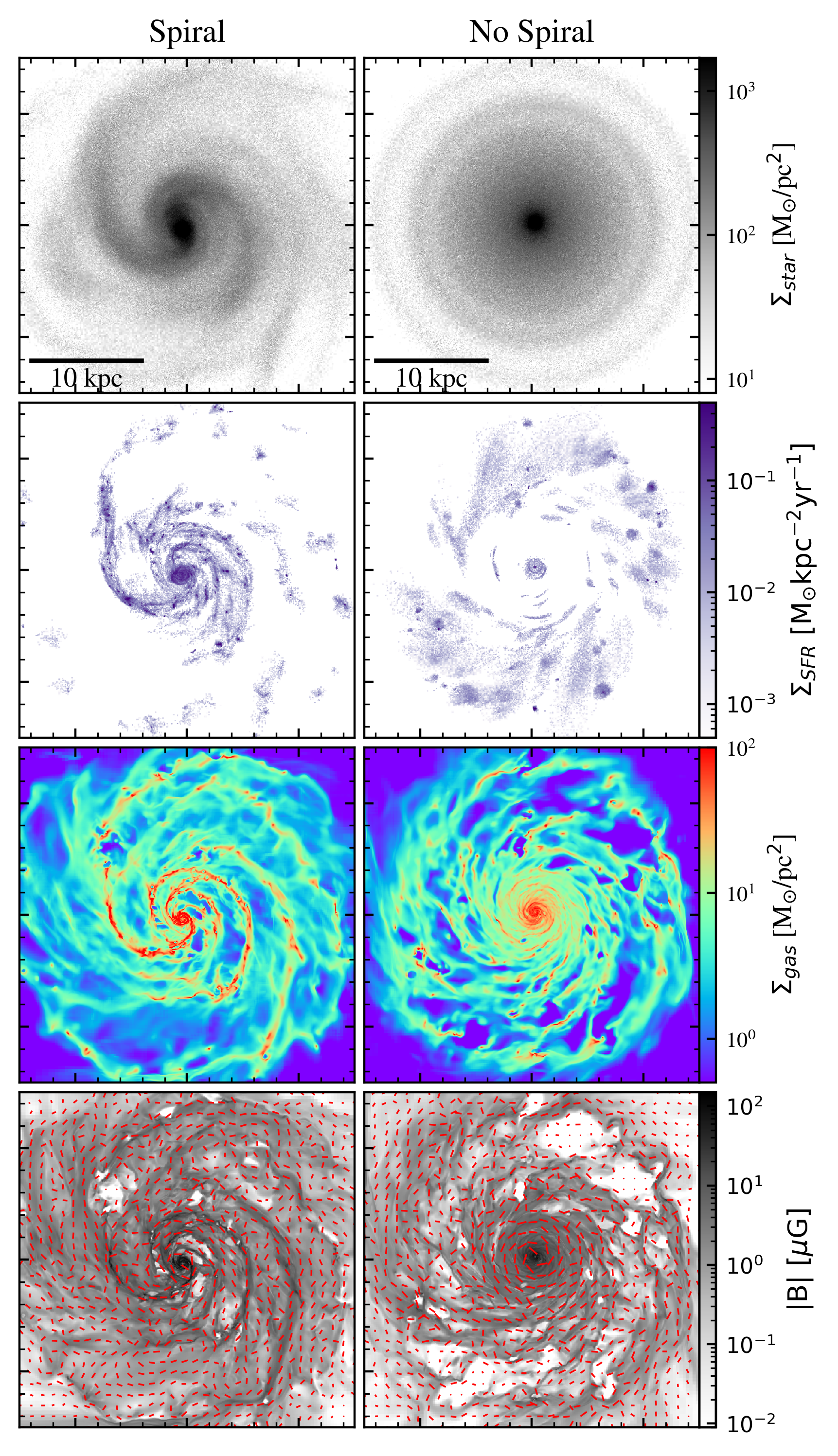}
\caption{Visualizations of stellar surface density (top row), star formation rate surface density (second row), gas surface density (third row), and mass-weighted magnetic field strength (bottom row) for each galaxy at t=600 Myr. Gas surface density and magnetic field strength are calculated using gas within 1 kpc of the midplane. SFR is calculated using all star particles that formed during the previous 100 Myr. The bottom row shows mass-weighted magnetic field strength smoothed over 1 kpc scales in grayscale with directions indicated by line segments.
\label{fig:visualisations}}
\end{figure}

\section{Simulation Results} \label{sec:results}
Both galaxies begin their evolution with an initial starburst that lasts $\sim$ 200 Myr. These initial bursts are typical in isolated galaxy simulations.  However, the amplitude is reduced here via a gradual ramping up of the star formation efficiency per free-fall time parameter. During the 1 Gyr of evolution, the spiral galaxy develops global spiral arms that are reflected in both the stellar and gas disks. The spiral arms are transient in nature, and shift between a 2, 3, and 4 armed spiral. The no-spiral galaxy remains axisymmetric in the stellar disk, but the gas disk develops some localized, spiral-like structure. These are associated with superbubbles that are sheared out into trailing shapes. After roughly 800 Myr, the spiral galaxy begins to develop a central bar in both the stars and gas. Interestingly, the bar does not form in the stars-only simulation (see Figure \ref{fig.proj}). 

Figure \ref{fig:visualisations} shows images of the stellar surface density, star formation rate surface density, gas surface density, and magnetic field strength after 600 Myr of evolution.  The stellar surface density in the upper left panel reveals a large-scale, mostly bi-symmetric, spiral pattern in the spiral galaxy.  The stellar disk in upper right panel shows the no-spiral model has remained axi-symmetric.

The spiral structure is also reflected in the star formation rate surface density (second row); almost all of the star formation occurs within the spiral arms of the spiral model, but is spread out diffusely in the no-spiral model.

The third row shows the surface density of gas in each model, created by a projecting of all gas within 1 kpc of the midplane. In the spiral model, gas is concentrated in the spiral arms, resulting in more high surface density gas in that case overall. In the no-spiral model, superbubbles carve large holes into the low-density ISM and create structures resembling higher multiplicity spirals as those regions are sheared out.
 
The bottom row shows a projection of mass-weighted magnetic field strength $|B|$, with lines indicating the direction of the magnetic fields within the plane of the disk. Mass-weighting favours the field in the denser gas tracing the spiral arms.
In both models, magnetic field strengths follow the density structure of the galaxy, with the interiors of superbubbles containing weak fields. The magnetic fields in the spiral galaxy are stronger, even in the inter-arm regions where the gas surface density is low.  In the spiral model, the mean field tends to follow the spiral arms in a manner similar to observed spiral galaxies \citep{fletcher2011}

\begin{figure}[ht!]
\includegraphics[width=\columnwidth]{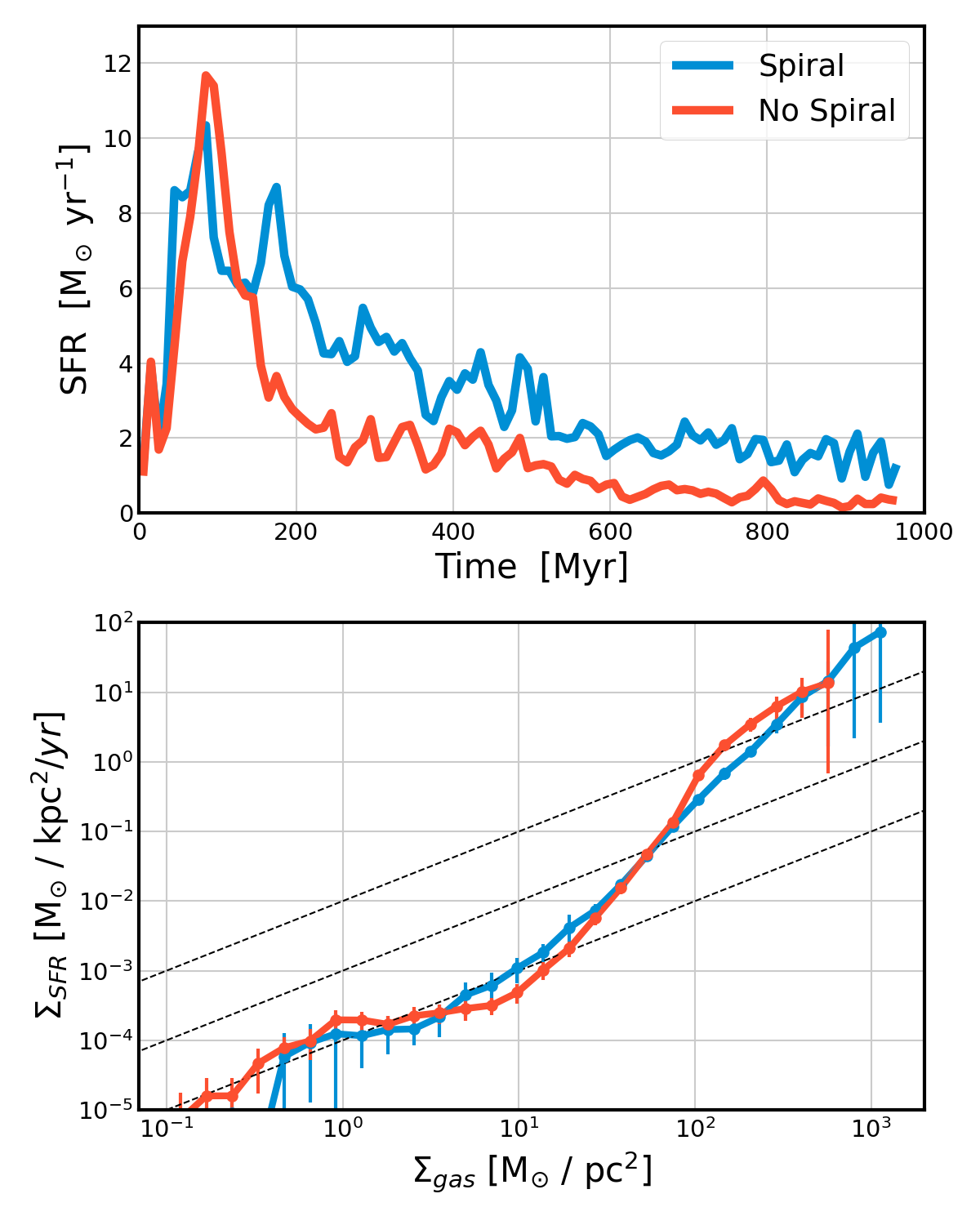}
\caption{Total SFR (top panel) and Kennicutt-Schmidt relation (bottom panel) for each galaxy. SFR is calculated by summing the mass of stars formed in each time bin and dividing by the bin size. Vertical lines show the 95\% confidence interval of the mean. Diagonal lines in the bottom panel indicate constant gas depletion times of 10$^8$, 10$^9$, and 10$^{10}$ years, as done in \citep{bigiel2008}.
\label{fig:sfr}}
\end{figure}

\begin{figure}[ht!]
\includegraphics[width=\columnwidth]{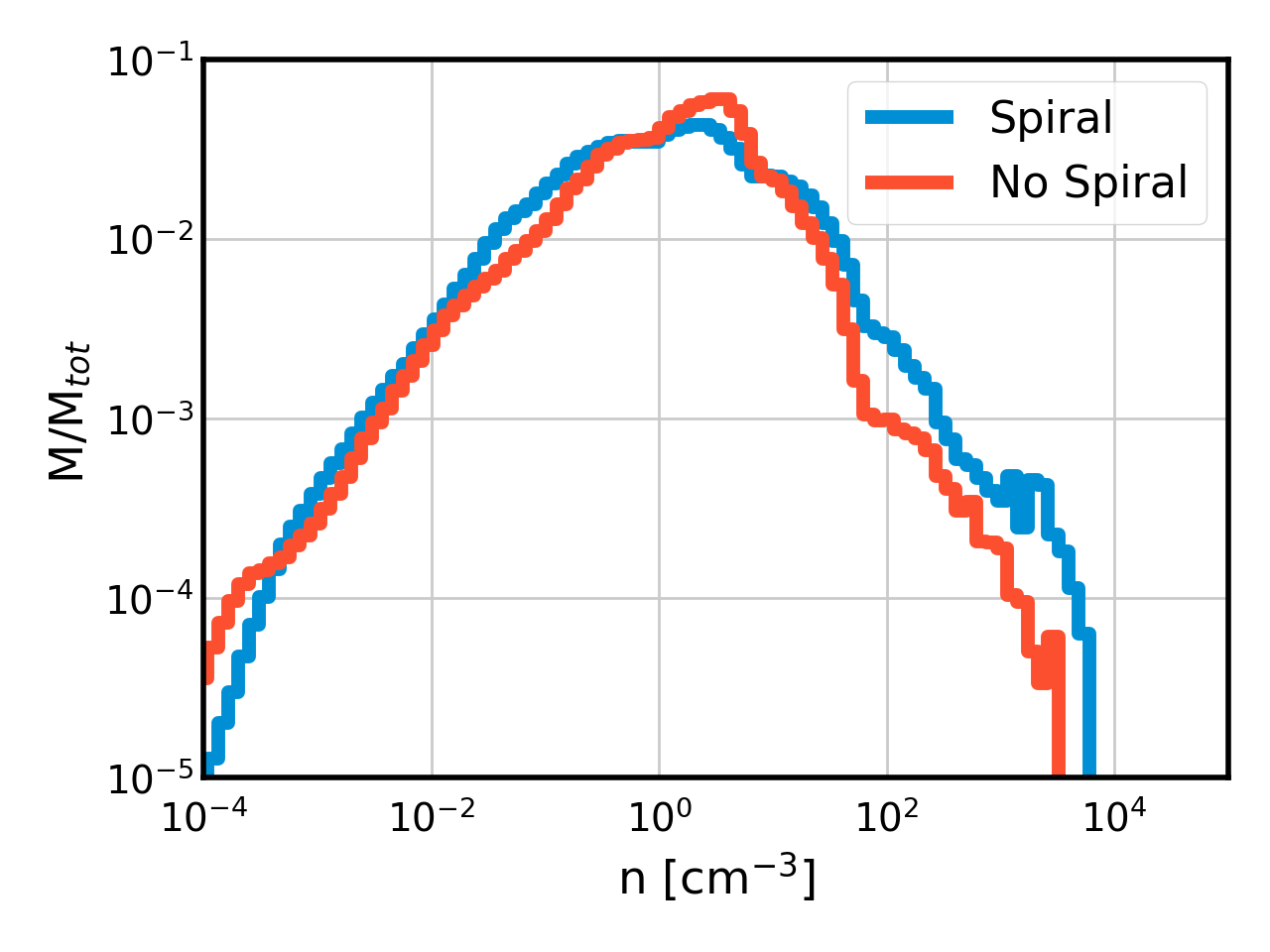}
\caption{Histogram of gas number densities. Each bin represents the total mass within that density range inside a disk with radius 15 kpc and 1 kpc height, normalized by the total mass within that disk. The spiral galaxy has more high density gas, which results in higher star formation rates.
\label{fig:density}}
\end{figure}

\subsection{Star Formation}
Figure \ref{fig:sfr} outlines the star formation histories of the galaxies. The top panel shows the time evolution of the total star formation rate. Both models begin with an initial burst of star formation that is followed by a gradual decrease.  After the initial burst, the spiral galaxy has a higher star formation rate (SFR) than the no-spiral galaxy for all of the subsequent evolution.  The star formation rate in the no-spiral galaxy drops to 0.3 M$_{\odot}$ yr$^{-1}$ after 1 Gyr of evolution, and the rate in the spiral galaxy is 1.2 M$_{\odot}$ yr$^{-1}$. This difference persists even though the no-spiral galaxy always has a larger reservoir of remaining gas; at 800 Myr there is 6.25 $\times 10^9$ M$_{\odot}$ of gas remaining in the spiral galaxy compared to the 7.43 $\times 10^9$ M$_{\odot}$ in the no-spiral galaxy. Between 300 and 800 Myr (after the initial starburst and before the bar forms in the spiral galaxy), the spiral galaxy has an average SFR 2.6 times higher than the no-spiral galaxy. This result alone shows that the presence of spiral arms in galaxies can cause a significant increase in the star formation rate.

The lower panel of figure \ref{fig:sfr} shows the Kennicutt-Schmidt relation for each model, which is calculated by dividing the galaxy into 100 pc sized patches and summing the total mass of stars formed within the last 10 Myrs. To ensure a large enough sample, the calculation is done on 30 outputs from 500-650 Myr, and the mean SFR surface density is plotted in each bin. In both models there is a scatter of over 6 orders of magnitude, but the mean SFR surface densities are generally quite similar at a given gas surface density. There are some minor differences.  At surface densities of 10 M$_{\odot}$/pc$^2$, the spiral model has a factor of 2 higher SFR, and at 200 M$_{\odot}$/pc$^2$,  the no-spiral model has a factor of 2 higher SFR.

To understand the SFRs, the 3D density must also be taken into account because SFR is based on the local density via a Schmidt-law. In figure \ref{fig:density}, we show gas density histograms. The spiral galaxy has more gas at densities above 100 cm$^{-3}$, which is the star formation threshold. This is required to enable the overall higher SFR.  As noted above, we do not see a higher star formation rate for a given gas column density in figure~\ref{fig:sfr} in the spiral case.  
This is slightly counterintuitive: given the significant contribution of the vertical gravity field from the stellar disk, one might expect the gas to have smaller scale heights within the spirals, meaning higher 3D densities and higher SFRs for the same gas column. However, turbulence injected by stellar feedback supports the gas against gravity so that the average midplane 3D density is moderately higher in the quieter no-spiral model. For example, at a surface density of 10 M$_{\odot}$/pc$^2$, the no-spiral model has a average midplane density of 0.067 M$_{\odot}$/pc$^3$, and the spiral model has an average midplane density of 0.036 M$_{\odot}$/pc$^3$ (higher scale heights). This is consistent with  the spiral galaxy having more turbulence in this gas, with an average velocity dispersion of 12 km/s vs 6.5 km/s in the no-spiral galaxy. 
The increased turbulence in the spiral galaxy also drives more clumping to enable the higher star formation rate here.  This illustrates how complicated it is to connect column densities with star formation rates even with a relatively simple star formation model.

Thus the differences in SFR are mainly driven by the overall amount of high-column density gas, of which there is much more in the spiral model, seen gathered along the spiral arms in figure~\ref{fig:visualisations}). The spirals sweep up gas and increase the surface density without decreasing the depletion time. This favours the argument that spiral arms simply {\it gather}  star-forming gas \citep{elmegreen1986,Querejeta2024,Sun2024}, rather than acting as a mechanism to trigger more star formation at a fixed gas column. The difference in star formation rates plays a role in the magnetic field amplification due to increased turbulence injected from the stellar feedback, powering the small-scale dynamo.

\begin{figure}[ht!]
\includegraphics[width=\columnwidth]{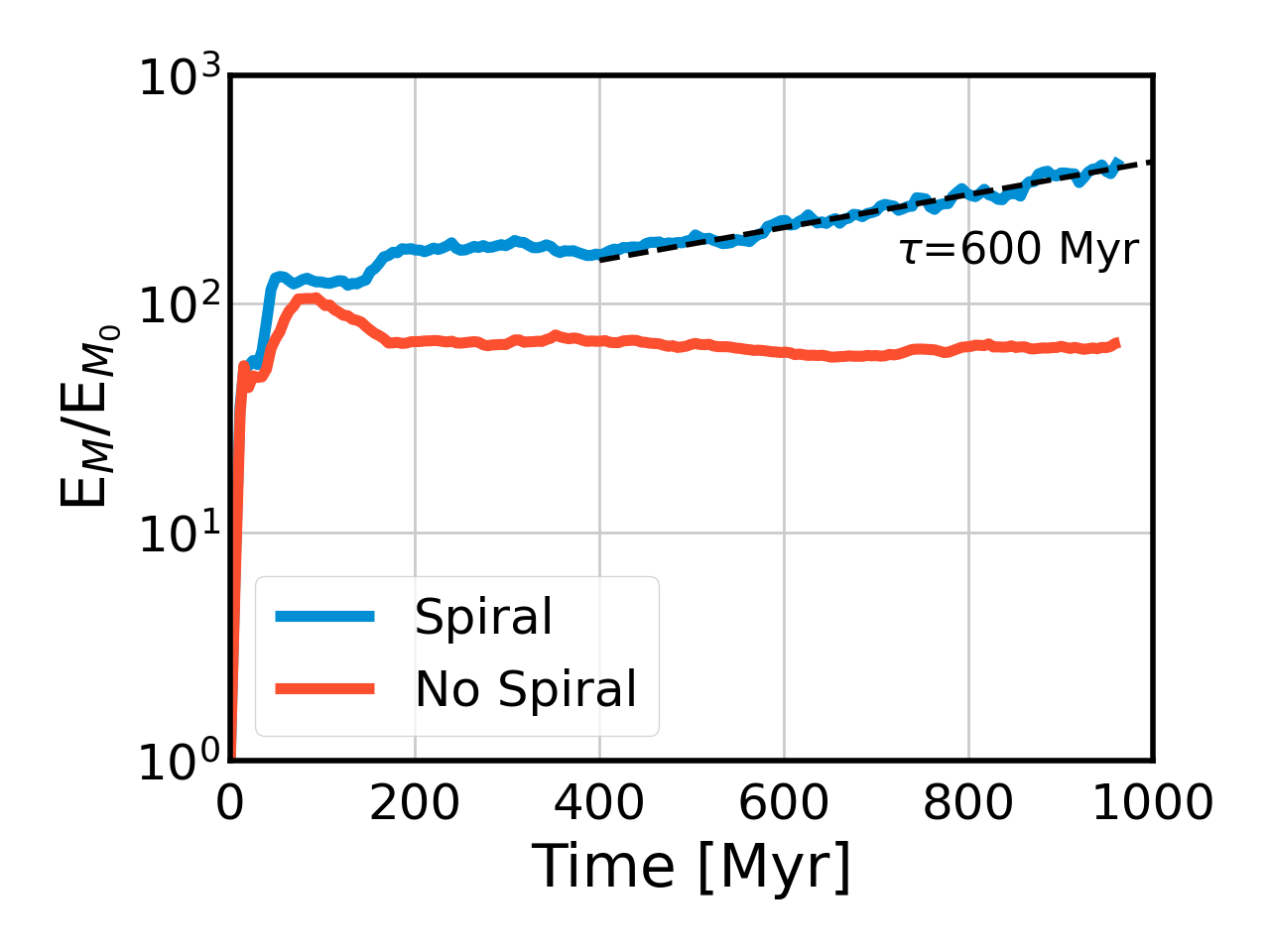}
\caption{Total magnetic energy evolution in each galaxy. Magnetic energy saturates in the no spiral galaxy after 200 Myr, but slower growth continues in the spiral galaxy. Dashed line shows exponential growth that goes as $E_M \propto$ exp(t/$\tau$).}
\label{fig:energy}
\end{figure}
\begin{figure}[ht!]
\includegraphics[width=\columnwidth]{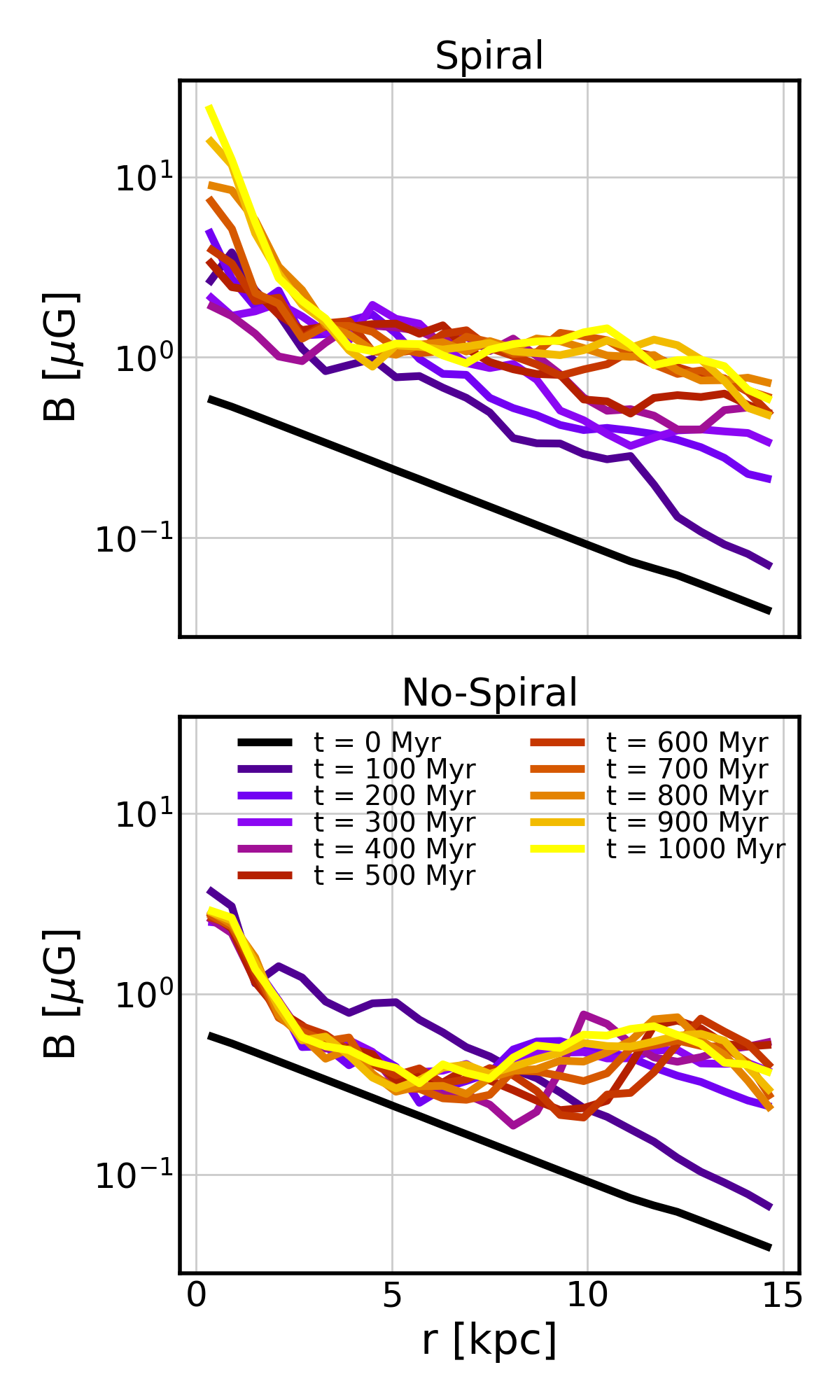}
\caption{Volume-averaged magnetic Field Strength versus radius over time in each galaxy. The spiral galaxy saturates with stronger fields than the no-spiral galaxy. 
\label{fig:B_vs_r}}
\end{figure}

\begin{figure*}[ht!]
\includegraphics[width=2\columnwidth]{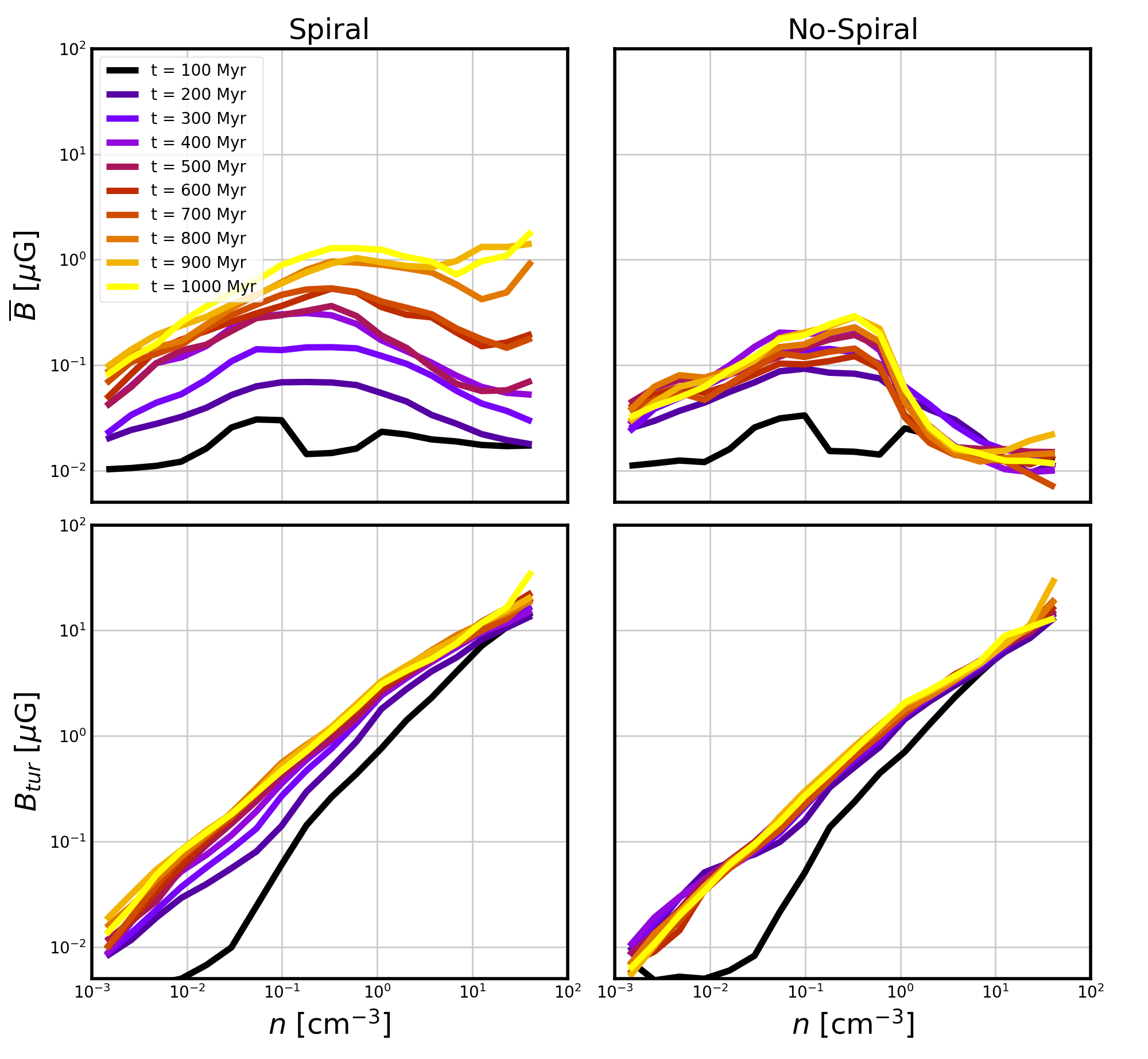}
\caption{Strength of the mean field (upper panels) and turbulent field (lower panels) versus density in each galaxy over time. Amplification of the mean field can be seen in the spiral galaxy in the upper-left panel. The mean field is calculated by applying a median filter over scales of 500 pc. Plotted values are are calculated within a disk of radius 15 kpc and height 1 kpc.
\label{fig:mean_field}}
\end{figure*}

\subsection{Magnetic Field Evolution}
\label{sec:field_evolution}
We now examine the evolution of the magnetic fields. Figure \ref{fig:energy} shows the evolution of the total magnetic energy within the whole simulation. In both galaxies, there is rapid growth during the first 100 Myr, coinciding with heightened turbulence associated with a moderate starburst.
This is consistent with the presence of an efficient small-scale dynamo \citep{Rieder1}. The spiral galaxy field grows due to second burst of star formation near $t$=180 Myr.  At late times, (500-1000 Myr), the spiral galaxy enters a phase of steady growth, with an exponential growth timescale of $\tau$ = 600 Myr. This is indicative of the presence of a slower, large-scale dynamo, as outlined in the following sections.

This growth can also be seen in figure~\ref{fig:B_vs_r}, which shows the volume-averaged magnetic field strength as a function of radius in each simulation over time.  For this analysis we choose to use volume-weighted fields as they reflect the total magnetic energy present and the support provided by magnetic pressure.  It also allows us to characterize the typical strength at a given radius as volume averages are less affected by the transient passage of density feature such as spirals.
Typical volume-averaged values are low ($\sim$ 1 $\mu$G) compared to mass-weighted spiral arm values ($\sim$ 3-30 $\mu$G) shown in figure~\ref{fig:visualisations}.  
Weighting choices were previously evaluated in \cite{Robinson2024}, who noted that
the strength ratio is fairly consistent between mass and volume weighting.  Thus volume weighting is a straight-forward way to examine the relative differences between the spiral and no-spiral cases.

Figure~\ref{fig:B_vs_r}  demonstrates how the magnetic fields amplify quickly during the first 200 Myr in both models, but they subsequently saturate outside $\sim 3$ kpc. Saturation occurs earlier in the no-spiral, while the spiral grows further to around 1 $\mu$G. At late times (600 Myr-1 Gyr), the resulting volume weighted average magnetic field strength in the spiral galaxy is 1 $\mu$G outside of the central 5 kpc. The no-spiral galaxy has weaker field strengths overall, with values ranging from 0.3-0.5 $\mu$G  past 5 kpc. In both models, the stronger fields correspond to regions of higher SFR. For example, in both models the fields are stronger in the center of the galaxy where star formation is concentrated, and in the no-spiral galaxy the field strengths increase in the outer disk where star formation is ongoing (see figure \ref{fig:visualisations}).  This shows that the small-scale dynamo plays an important role in setting the overall magnetic field strengths in both cases.

\subsection{Large-scale vs. Small-scale Field Evolution}

To further examine the large-scale dynamo, we decompose the magnetic field into mean and turbulent components, defined as $\mathbf{B_{turb}} = \mathbf{B}-\mathbf{\bar{B}}$, where the mean field $\mathbf{\bar{B}}$ is calculated using a median filter applied over 500 pc. Figure \ref{fig:mean_field} shows the mean and turbulent field vs. gas number density in each galaxy over time. 

The upper left panel shows that the large-scale dynamo in the spiral galaxy has continuously enhanced the mean field at all densities. At 1 Gyr, intermediate density gas has reached a mean field strength of 1 $\mu$G. The upper right panel shows the mean field in the no-spiral galaxy, which has limited long-term amplification. We note that $\sim$1 cm$^{-3}$ marks a transition from diffuse to dense gas structures. 

The bottom panels show the turbulent component, which is stronger above $\sim0.3$ cm$^{-3}$ in both galaxies. The turbulent field follows similar saturation behaviour to the total field shown in figure~\ref{fig:B_vs_r}, saturating earlier in the no-spiral at a lower value for a given density.  It remains saturated in both cases from $\sim 500$ Myr.  The higher turbulent field at saturation in the spiral case is expected due to the higher levels of turbulence and feedback in that case. 

The turbulent field retains power-law behaviour similar to the 2/3 power law that was included in the initial condition and flattening to a roughly 1/2 power law at moderate to higher densities. A power-law relationship between field strength and gas density, $B \propto \rho^\alpha$, is seen in both observations and in a wide range of MHD simulations of processes in the interstellar medium \citep{whitworth2025}.  It is found to vary typically between $1/2$ and $2/3$ .  Field growth at fixed density, as shown in figure~\ref{fig:mean_field}, is a clear signal of dynamo action.

\begin{figure}[ht!]
\includegraphics[width=\columnwidth]{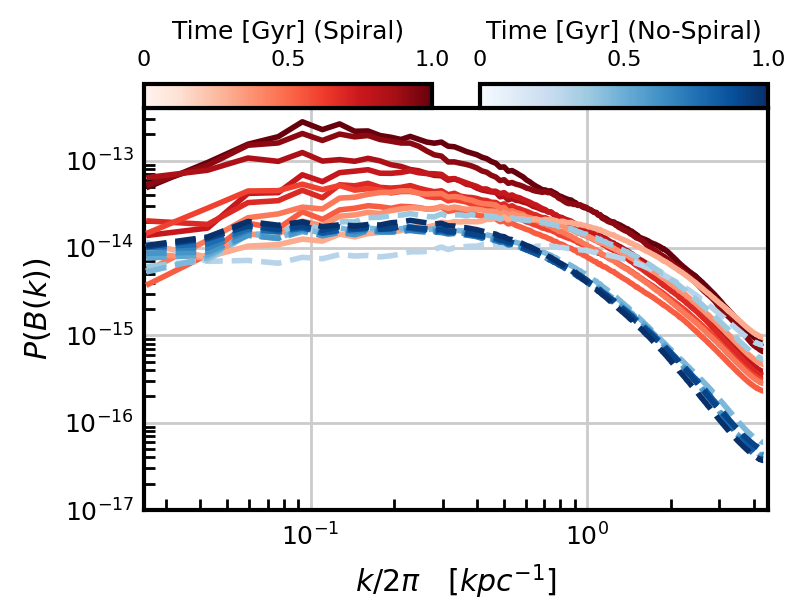}
\caption{Total magnetic field power spectra in each galaxy. Each line shows a different time, going from 100 Myr (light shaded) to 1 Gyr (dark shaded). Power spectra are calculated in a cube of side-length 9.375 kpc centered on the galaxy. The power spectrum in the spiral galaxy shows continued growth up to 1 Gyr, while in the no-spiral galaxy there is no long-term growth. The power spectrum in the spiral galaxy peaks at scales near 10 kpc, which is a similar scale to those of spiral arms. 
\label{fig:power_spectra}}
\end{figure}

\begin{figure*}[ht!]
 \includegraphics[width=2\columnwidth]{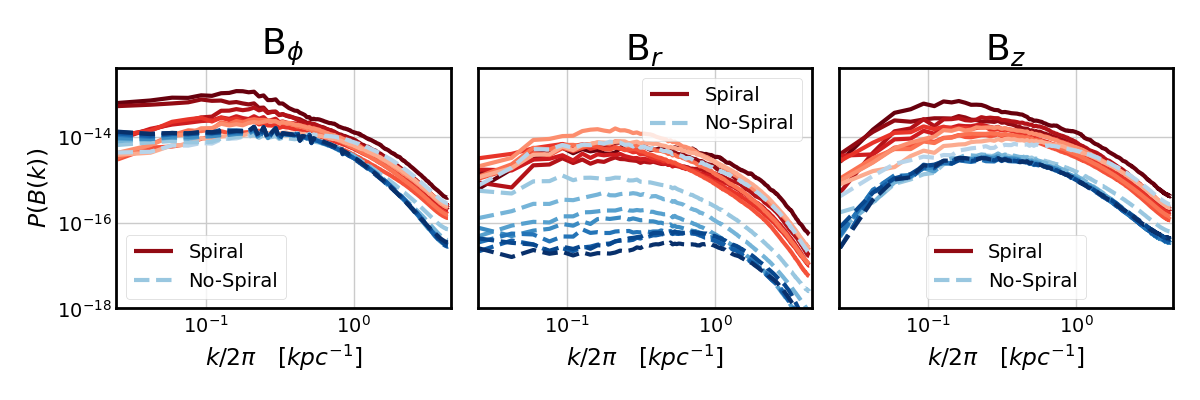}
\caption{Power spectra of the azimuthal, radial, and vertical field components over time. Each line shows a different time, going from 100 Myr (light shaded) to 1 Gyr (dark shaded).
\label{fig:power_toroidal}}
\end{figure*}

\subsection{Magnetic Field Power Spectra}

To better understand the scale dependence of the magnetic field evolution, we calculate the power spectra of the magnetic field and its cylindrical components. In figure \ref{fig:power_spectra}, we plot the total angle-averaged power spectrum of the magnetic field defined as 
\begin{equation}
    P(k) = \langle\tilde{B}(k) \cdot \tilde{B}(k)^*\rangle
\end{equation}
where $\tilde{B}$ is the Fourier transform of the magnetic field, $k$ is the wavevector, and * indicates the complex conjugate. The angled brackets indicate an 'angle-average' which yields a one-dimensional power spectrum by binning $P(k)$ in wavevectors $k=|k|$ and dividing by the number of points in each bin \citep{Joung2006}. The power spectra are calculated on an 18.3 pc resolution cube of side-length 10 kpc centered on the galaxy.

In figure \ref{fig:power_spectra}, the magnetic field continues to grow on all scales in the spiral galaxy throughout the 1 Gyr of evolution, compared to the no-spiral model which shows little to no growth at all scales. This reflects the stronger magnetic fields in the spiral galaxy discussed in the previous section spiral galaxy. However, at late times there is more growth in the spiral galaxy on large ($\gtrsim$ 10 kpc) scales,  which implies a large scale ordering of the field on the scale of spiral arms. 



Figure \ref{fig:power_toroidal} breaks the magnetic field power spectrum  into its three cylindrical components: vertical, radial, and azimuthal. In the left panel, the toroidal component can be seen steadily growing in the spiral galaxy, but in the no-spiral galaxy there is little growth. In the radial component, the power spectrum of the spiral galaxy remains steady, but the no-spiral galaxy consistently loses power at all scales over time. The z-component steadily grows in the spiral galaxies, and slightly loses power in the no-spiral galaxy.

After 1 Gyr of evolution, the field is dominated by the azimuthal component in both galaxies. The vertical ($z$) component is a factor of two weaker than the toroidal at 10 kpc scales in the spiral galaxy, and a factor of five less in the no-spiral galaxy. 
 Power in the z-component drops steeply at large scales, which implies that the largest-scale vertical field is created by kpc-scale feedback events that drag magnetic fields upwards.  These events are galactic fountains with material reaching heights of 3-5 kpc, which is consistent with the peak of the z-component power spectrum. The radial field component is the weakest, being almost two orders of magnitude lower than the toroidal component in the spiral galaxy, and almost 3 orders of magnitude lower in the no-spiral galaxy.

\begin{figure}[ht!]
\includegraphics[width=\columnwidth]{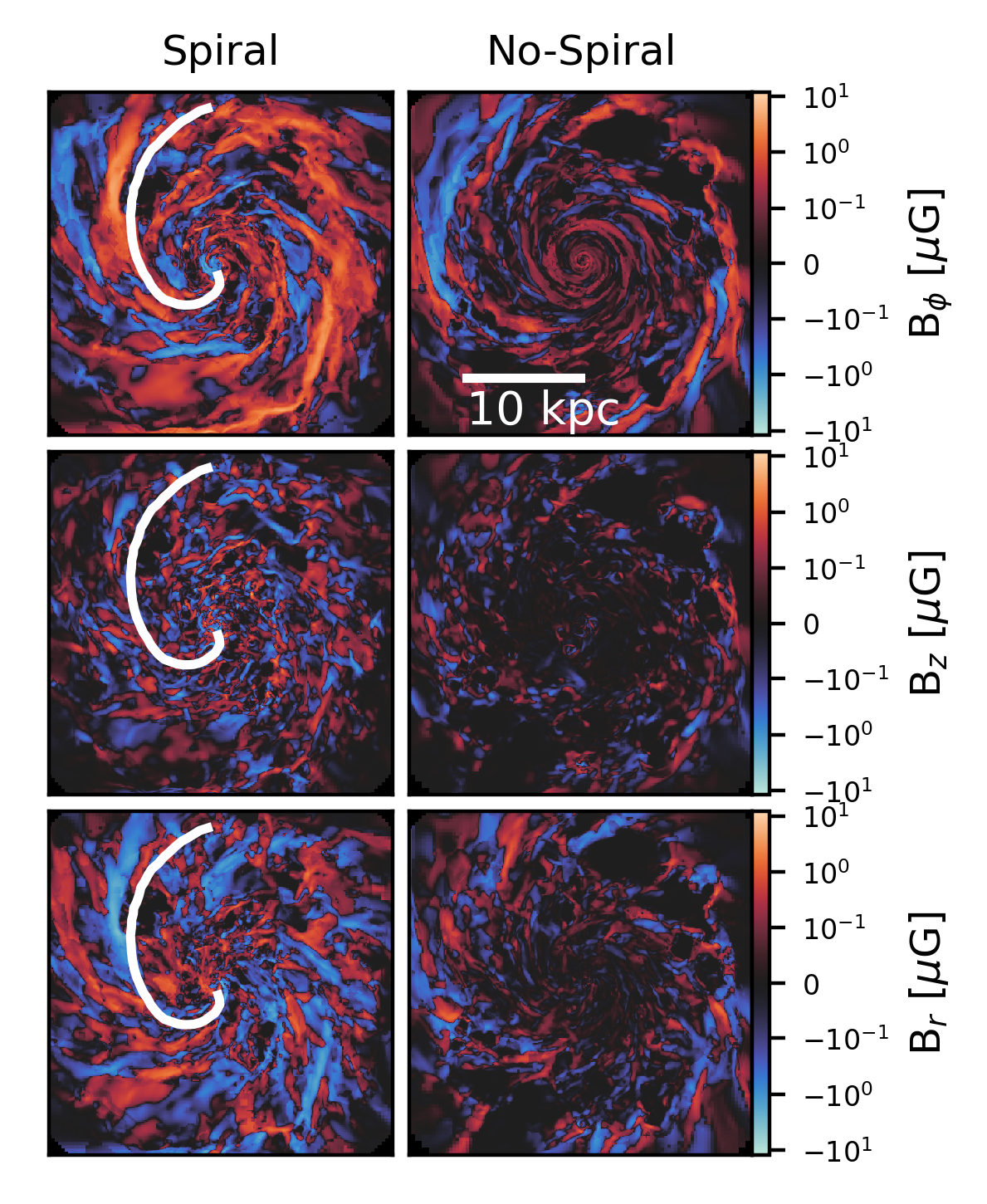}
\caption{Projections of azimuthal (top row), vertical (middle row), and radial (bottom row) magnetic field components, at time t = 600 Myr. The white solid line shows the location of the left spiral arm in the stellar surface density shown in fig.~\ref{fig:visualisations}.
\label{fig:field_components}}
\end{figure}

The components of the field are visualized in figure \ref{fig:field_components}. In the top row, the toroidal component of the field shows ordered fields in the spiral galaxy. In the upper left panel, there is a coherent field with strengths near 10 $\mu$G  following the spiral arm structure, disrupted only by the formation of a superbubble, and there are field reversals in the inter-arm regions. In the no-spiral galaxy,  there is no spiral structure to follow and the field becomes increasingly wound up in the central 10 kpc, which is a region that has low star formation rates. The z-component of the field is more randomly distributed in both galaxies, with no visible structure linked to the spiral arms. 
This supports the idea that the vertical component of the field is associated with localized turbulence. The overall strength of the vertical field is weaker in the no-spiral galaxy, because of the lower star formation rate sustains weaker feedback-driven turbulence. The radial components of the field are shown in the bottom row, which also shows large scale structure associated with the spiral arms, which was reflected in the power spectrum.

\begin{figure}[ht!]
\includegraphics[width=\columnwidth]{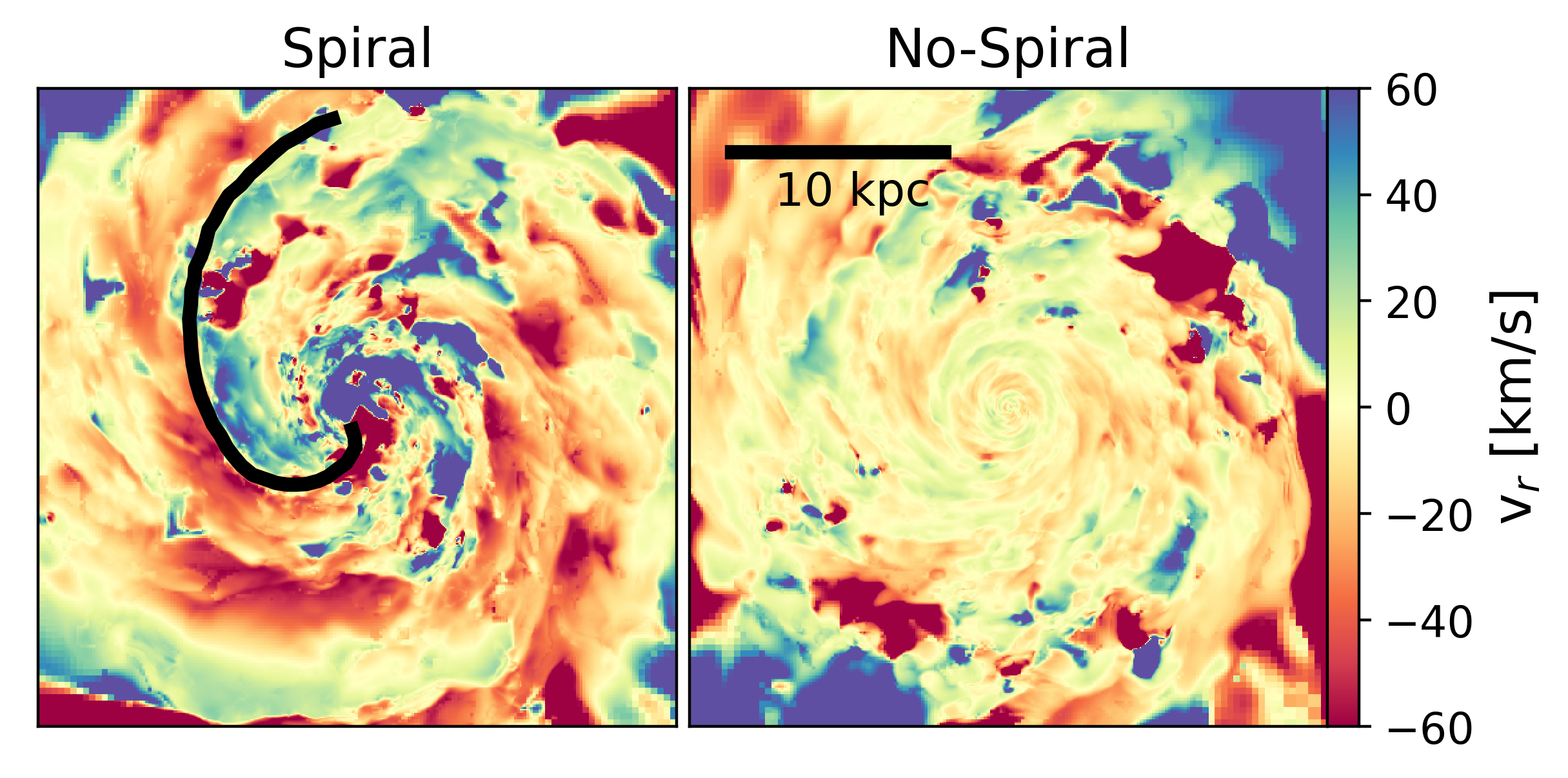}
\caption{Mass-weighted vertical average of the radial velocity of gas inside a disk of 1 kpc height at t= 600 Myr. Radial velocities are in the plane of the disk, and are positive for outward motion. The solid line shows the location of a spiral arm identified in the stellar surface density shown in fig.~\ref{fig:visualisations}. 
\label{fig:vr}}
\end{figure}

\subsection{Dynamo Mechanism}
\label{sec:mechanism}

The field strengths in figure \ref{fig:B_vs_r} 
and the bottom panel of figure ~\ref{fig:mean_field}  provide convincing evidence for a small-scale dynamo powered by turbulent motions, which were in turn generated via feedback from star formation.
 
In the earliest phases, the fields are not strong enough to affect the gas flows, so the small-scale dynamo operates in the kinetic regime which leads to exponential growth rates. The small-scale dynamo saturates when the fields become strong enough to back-react onto the flow. Rapid field growth followed by saturation or slower growth rates is consistent with the two phases of amplification demonstrated in simulations by \cite{Rieder1, Rieder2}.

In addition to the small-scale dynamo, a mean-field dynamo similar to that proposed by \cite{sellwood2002} can explain the mean-field amplification and power spectra.  In this mechanism, the dynamo is powered by radial flows driven by the spiral arms. 

This process will generate large-scale radial fields as the primarily toroidal fields are dragged radially by large-scale gas flows near spiral arms. In the no-spiral galaxy there are no coherent motions to regenerate these radial fields, and the radial field components continually decay as they are sheared into toroidal fields by the $\Omega$ effect. In essence, the mean-field growth is limited because large scale radial fields have no generation mechanism. In the spiral galaxy, radial fields are also sheared out by the $\Omega$ effect, however the spiral arms provide a mechanism that can constantly replenish the radial fields at the same time they are being sheared out.

Figure \ref{fig:vr} shows the radial velocity in each simulation at 600 Myr, with the location of one spiral arm shown by a solid line.  The spiral is created by the streaming flows of stars and gas around the galaxy converging as they approach the arm and diverging again after passing through it.  The motion is predominantly azimuthal, with circular velocities near 200 km/s through most of the disk, with the spiral potential adding a coherent radial part, with speeds up to $\sim50$ km/s that cause the stream lines to converge and diverge at the appropriate phases.  These flows are completely absent in the no-spiral galaxy due to the absence of spiral arms.  In the spiral galaxy, the small but coherent radial flows will drag magnetic fields with them, constantly replenishing the radial fields, which is a key process in maintaining the mean-field dynamo. 

We can compare these results with established mean-field dynamo theories. Classical $\alpha-\Omega$ dynamo theory predicts a quadrupolar field being generated in the galactic halo \citep{stix1975,shukurov2019}, which has been demonstrated in simulations by \cite{ntormousi2020}. We do not see evidence for quadrupolar fields in the galactic halo of our simulations, nor the presence of individual Parker loops twisting above the spirals. 
The material 3-5 kpc above the disk is associated with galactic fountains from local feedback events. 
However, because of the lower numerical resolution in the halo due to the refinement scheme used in \textsc{RAMSES}, we cannot make strong claims about behaviour in the galactic halo. Another factor is that our simulations were only evolved for 1 Gyr which may not be long enough for the  $\alpha-\Omega$ dynamo to fully develop. A clear similarity to the  $\alpha-\Omega$ dynamo that we have shown is a conversion of fields between radial and toroidal components, resulting in a net-increase of toroidal field strength.

Despite the strong turbulent fields in high-density gas, the majority of magnetic energy resides in the mean field, due to the low density gas having a much higher volume. This means that the mean-field plays a dominant role in the large-scale dynamo amplification from 600-1000 Myr. The radial fields that are used in the dynamo are the large-scale mean-field created by the spiral flows, rather than turbulent fields created by local feedback events

Our simulations manifest spiral driven radial flows that we identify as promoting the large-scale dynamo, but they lack the resolution to fully capture the dynamics of a realistically clumpy ISM.  The radial migration of gas clouds in response to transient spirals, predicted by \citep{sellwood2002} and described here in the introduction, would provide a further driver of large-scale turbulence, and perhaps lead to more vigorous growth of large scale mean field.  
 
\section{Discussion} \label{sec:discussion}

This work shows that star formation and magnetic field amplification are qualitatively different when spiral arms are included. However, there are several caveats. 



We primarily employ volume-weighted field measurements (typical values $\sim 1$ $\mu$G) for the reasons discussed in section~\ref{sec:field_evolution}.  Mass-weighted measures are typically higher, in the range $3-30$ $\mu$G (as shown in figure~\ref{fig:visualisations}).  Volume-weighted averages of $3-30$ $\mu$G can also be inferred for denser gas $> 1$ cm$^{-3}$, as shown in figure~\ref{fig:mean_field}.
For comparison, large-scale field strengths of order 10-20 $\mu$G commonly reported in observations of nearby spiral galaxies \citep{fletcher2011,basu,Beck}.  Those values reflect specific emission processes and assumptions that were made to interpret them and do not correspond directly to either of our weighting scheme \citep{basu,ponnada_2022}.  Synchrotron observations, in particular, often assume an equipartition of energy between magnetic fields and cosmic rays, which may lead to the magnetic field strengths being overestimated \citep{Dacunha2025}. 

Additionally, the simulations do not resolve high-density gas which has the strongest fields.  This is the gas probed with Zeeman effect measurements \citep{crutcher,whitworth2025}.  It would be interesting to explore detailed mock-observations in future work.



This work also does not account for the evolution of cosmic rays or non-ideal MHD effects like ambipolar diffusion. Ambipolar diffusion operates on scales that are small compared to these simulations so it is unlikely to be important. Cosmic rays are closely associated with magnetic field evolution and affect overall gas support \citep{semenov2021}. There is still considerable uncertainty in modeling cosmic rays, but including them in these simulations is another future direction.

Our result that the generated fields are mainly toroidal may be influenced by the initial condition which contains a weak toroidal field. Toroidal fields are the natural choice because any initial radial or vertical components would rapidly be sheared out into a toroidal field. Nonetheless, it would be an interesting study to begin with several different field morphologies, and determine if the results remain the same. We also do not include more advanced stellar feedback models such as the superbubble model \citep{keller2014}, which may drive turbulence more effectively and effect the small-scale dynamo. Stronger feedback may also drive larger-scale outflows and galactic fountains, which could generate large-scale z-component fields, or affect the development of quadrupolar fields in the halo.

\section{Conclusions} \label{sec:conclusions}
In this work, we simulated two magnetized disk galaxies that are manifestly the same except for the presence or absence of stellar spiral arms. This is a novel addition to the recent literature concerning global MHD galaxy simulations \citep[e.g. ][]{Rieder2,butsky2017,pakmor,ntormousi2020,steinwandel2,ponnada_2022,Liu2022,wissing,wibking2023,bo2024,martin2024}.  

We follow the methods of \cite{Robinson2024} and examine the effects of spiral arms on both magnetic field amplification and star formation. The spiral arms were allowed to naturally evolve in the first galaxy, and in the second, a fixed gravitational potential acted on the star particles to keep the stellar disk axisymmetric. 
The spiral galaxy has higher star formation, driving a stronger small-scale dynamo.  A slower, large-scale dynamo is linked to gas flows along the spiral arms.  

We summarize our conclusions as follows:

\begin{itemize}
\item   Spiral arms promote star formation.  The simulated spiral galaxy maintains a 2.6 times higher SFR.  This is due to high column gas being swept up by the spiral arms (even though there is less gas remaining overall). The star formation rate efficiency remains similar at a given gas column, which supports the conceptual picture that spiral arms {\it gather} star formation.

\item The turbulent component of the magnetic field dominates at early times ($<200$ Myr) and remains stronger at higher densities.  The magnetic field energy tracks the star formation rate and associated turbulence for the first $\sim 200$ Myr. The spiral galaxy generates early field strengths that are a factor of 2-3 stronger in the disk than in the no spiral case due to higher star formation.

\item The small-scale dynamo generates field in all directional components.  However, radial fields are converted to toroidal fields by shear, which produces a long-lived toroidal field.  This progressively weakens the radial field in the no-spiral case.


\item In the spiral galaxy, the mean-field is amplified by a large-scale dynamo. The mean field strength rises to 1 $\mu$G (volume-averaged) over the Gyr of evolution, with an e-folding time of 600 Myr.  The no-spiral galaxy mean field saturates at $\sim 0.2$ $\mu$G after 400  Myr, and has no ongoing amplification.

\item The spiral large-scale dynamo seen here operates by converting toroidal fields into radial fields and back again.  Coherent, large-scale radial gas flows associated with the spiral arms drag the toroidal field to create radial field.  Net additional toroidal field is generated after shearing out the radial field.  No large-scale vertical field is created.  

\end{itemize}

\section*{Acknowledgements}
HR is supported by an NSERC postgraduate scholarship, and 
JW and REP are supported by Discovery Grants from NSERC of Canada.  Computational resources for this project were enabled by a grant to JW from Compute Canada/Digital Alliance Canada and carried out on the Niagara Supercomputer.
JAS acknowledges the continuing hospitality and support of Steward Observatory.

\def\etal{{\it et al.}}
\def\ie{{\it i.e.}}
\long\def\Ignore#1{\relax}

\appendix

\section{}
Our initial disk+bulge+halo model is, like the AGORA model
\citep{Kim2014}, designed to resemble an axisymmetric Milky Way.  It
differs from their model by having a more massive disk and bulge, in
order that it will support fewer spiral arms, while the dense bulge
component makes the model less prone to bar instabilities.  The
equilbrium model has three massive components that are all realized
with collisionless particles.  Our simulations with the Ramses code
employ a smooth gas component that had a mass of 10\% of the disk.  We
therefore reduced the masses of each disk particle by 10\% in order
to maintain gravitational equilibrium.  Alternatively, if the user
prefers a code that employs gas particles, the appropriate particle
fraction may be labeled by the user as gas particles as desired.

\subsection{Disk}
The disk is a thickened exponential with radial scale length $R_d$
having the volume density
\begin{equation}
\rho_d(R,z) = {M_d \over (2\pi)^{3/2} R_d^2}e^{-R/R_d} \exp\left(
    {-z\over 2z_0}\right)^2,
\end{equation}
with $z_0$ being the constant disk scale height.

\subsection{Bulge}
The spherical bulge has the cusped density profile proposed by
\citet{Hern90}
\begin{equation}
\rho(r) = {M_ba_b \over 2\pi r(r+a_b)^3},
\end{equation}
with $M_b$ being the bulge mass and $a_b$ a scale radius.  Hernquist
also supplied an isotropic distribution function (DF) for this isolated
mass distribution.

\subsection{Halo}
We start from a second, more extensive and massive spherical Hernquist
model for the halo component, having mass $M_h$ and radial scale
$a_h$, and impose an outer boundary at $r=8a_h$ to the otherwise
infinite halo.  Since the central attraction of the disk and bulge,
which are embedded at the center of the halo, destroy the radial
balance of the isotropic DF supplied by Hernquist, we must derive a
revised equilibrium DF for the halo of the composite model.

Our method to achieve this is a follows: We start from a known DF for
the halo with no embedded disk or bulge and compute its density change
assuming that the masses of the disk and bulge were increased
adiabatically from zero.  As was pointed out by \citet{Young80},
adiabatic changes to the total mass profile can be calculated
semi-analytically, since the actions \citep{binney_tremaine} of the halo
particles do not change as the potential well changes slowly;
therefore the DF expressed as actions is the same after the adiabatic
change as before.

\citet{Blum86} assumed angular momentum alone, one of the
actions of an orbit, was conserved, but their formula would apply only
if the orbits of all particles were initially and remained precisely
circular.  The orbits of particles in all reasonable spherical models
librate radially, and therefore one must take conservation of radial
action into account when computing the density response to adiabatic
changes to the potential, and the pressure of radial motions makes a
realistic halo more resistant to compression than the na\"{\i}ve model
with no radial action would predict.

\citet{Young80} and \citet{SM05} describe procedures, which we employ
here, to include radial action conservation as the potential well
changes adiabatically.  An initially isotropic DF becomes mildly
radially biased, which can still be represented by the unchanged
actions.  Note that the procedure assumes the potential remains
spherically symmetric, so we must approximate the disk potential by
the monpole only term, \ie\ using only the disk mass enclosed in a
sphere of radius $r$.  \citet{SM05} found, from a comparison with a
simulation in which a disk was grown slowly inside a spherical halo
that the ashperical part of the disk potential caused negligibly small
changes to the spherically averaged halo potential.

\begin{figure}
\begin{center}
\includegraphics[width=.99\hsize,angle=0]{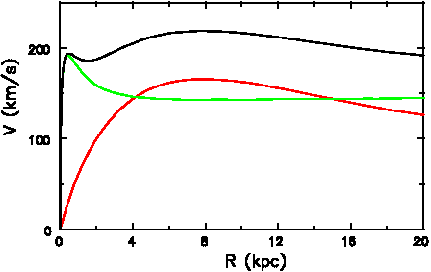}
\end{center}
\label{fig:rotation_curve}
\caption{The black line gives the total rotation curve of our model,
  after halo compression, with the separate contributions of the
  thickened disk (red line) and spherical bulge (green line).}
\label{fig.MWRC}
\end{figure}

\begin{table}
\caption{Our Milky Way model}
\label{tab.pars}
\begin{tabular}{@{}lll}
  Disk & mass & $M_d = 6.10 \times 10^{10}\;$M$_\odot$  \\
            & scalelength & $R_d = 3.5\;$kpc  \\
            & vertical thickness & $z_0 = 350\;$pc \\
            & in-plane motion & $Q=1.2$ \\
            & gravity softening & $\epsilon = 70\;$pc \\
  Bulge &  mass & $M_b = 4.07 \times 10^{10}\;$M$_\odot$  \\
           & scale radius & $a_b = 0.35\;$kpc  \\
  Initial halo & mass & $M_h = 73.21 \times 10^{10}\;$M$_\odot$  \\
           & scale radius & $a_h = 52.5\;$kpc  \\
\end{tabular}
\end{table}

\begin{figure*}
\begin{center}
\includegraphics[width=.7\hsize,angle=270]{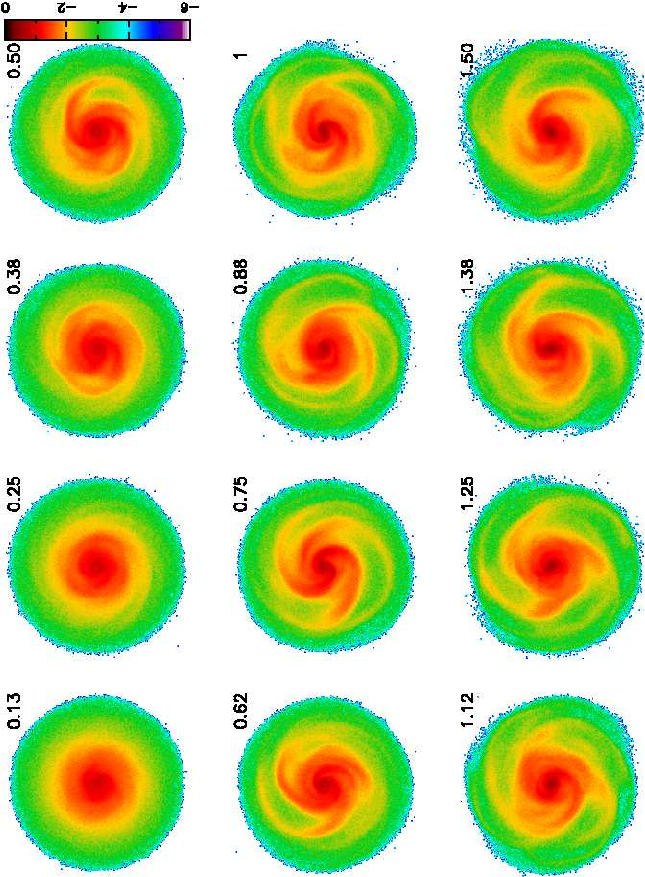}
\end{center}
\caption{The first 1.5 Gyr of evolution of a model having no gas
  particles.  The outer radius of the disk in 20~kpc.  Notice the
  strong, open spirals and the absence of a large bar.}
\label{fig.proj}
\end{figure*}

With the compressed halo potential and DF in hand, we have the global
potential of the thickened disk, bulge and halo, and can set the
orbital motions of the particles in the usual way \citet{Sell24}.
Hernquist's isotropic DF adequately describes the bulge equilibrium,
since that component dominates the potential in the center.

\subsection{Selection of parameters and scaling}
Our choices for the parameters of the model are given in
Table~\ref{tab.pars} which, after halo compression result in the
rotation curve shown in Fig.~\ref{fig.MWRC}.  \Ignore{We have adopted
a scaling such that the unit of time is 12.5\;Myr.}

We can realize this model with any number of particles in each
component, and find that it is very close to equilibrium with $T/|W|
\simeq 0.504$.  Fig.~\ref{fig.proj} illustrates the first 1.5~Gyr of
evolution of the disk, which is modeled by 1M star particles and no
gas component.  The bulge and halo are not shown but are represented
by 0.1M and 0.5M particles respectively.











\bibliography{example}{}

\begin{thebibliography}{}
\expandafter\ifx\csname natexlab\endcsname\relax\def\natexlab#1{#1}\fi
\providecommand{\url}[1]{\href{#1}{#1}}
\providecommand{\dodoi}[1]{doi:~\href{http://doi.org/#1}{\nolinkurl{#1}}}
\providecommand{\doeprint}[1]{\href{http://ascl.net/#1}{\nolinkurl{http://ascl.net/#1}}}
\providecommand{\doarXiv}[1]{\href{https://arxiv.org/abs/#1}{\nolinkurl{https://arxiv.org/abs/#1}}}

\bibitem[{{Agertz} {et~al.}(2011){Agertz}, {Teyssier}, \& {Moore}}]{delayed_cooling}
{Agertz}, O., {Teyssier}, R., \& {Moore}, B. 2011, \mnras, 410, 1391, \dodoi{10.1111/j.1365-2966.2010.17530.x}

\bibitem[{{Balbus} \& {Hawley}(1991)}]{balbus1991}
{Balbus}, S.~A., \& {Hawley}, J.~F. 1991, \apj, 376, 214, \dodoi{10.1086/170270}

\bibitem[{{Basu} \& {Roy}(2013)}]{basu}
{Basu}, A., \& {Roy}, S. 2013, \mnras, 433, 1675, \dodoi{10.1093/mnras/stt845}

\bibitem[{{Beck} {et~al.}(2020){Beck}, {Berkhuijsen}, {Gie{\ss}{\"u}bel}, \& {Mulcahy}}]{Beck2020}
{Beck}, R., {Berkhuijsen}, E.~M., {Gie{\ss}{\"u}bel}, R., \& {Mulcahy}, D.~D. 2020, \aap, 633, A5, \dodoi{10.1051/0004-6361/201936481}

\bibitem[{{Beck} {et~al.}(2019){Beck}, {Chamandy}, {Elson}, \& {Blackman}}]{Beck2019}
{Beck}, R., {Chamandy}, L., {Elson}, E., \& {Blackman}, E.~G. 2019, Galaxies, 8, 4, \dodoi{10.3390/galaxies8010004}

\bibitem[{{Beck} \& {Wielebinski}(2013)}]{Beck}
{Beck}, R., \& {Wielebinski}, R. 2013, in Planets, Stars and Stellar Systems. Volume 5: Galactic Structure and Stellar Populations, ed. T.~D. {Oswalt} \& G.~{Gilmore}, Vol.~5, 641, \dodoi{10.1007/978-94-007-5612-0_13}

\bibitem[{{Bell}(2004)}]{bell2004}
{Bell}, A.~R. 2004, \mnras, 353, 550, \dodoi{10.1111/j.1365-2966.2004.08097.x}

\bibitem[{{Berrier} \& {Sellwood}(2015)}]{Berrier2015}
{Berrier}, J.~C., \& {Sellwood}, J.~A. 2015, \apj, 799, 213, \dodoi{10.1088/0004-637X/799/2/213}

\bibitem[{{Bhat} {et~al.}(2016){Bhat}, {Subramanian}, \& {Brandenburg}}]{Bhat2016}
{Bhat}, P., {Subramanian}, K., \& {Brandenburg}, A. 2016, \mnras, 461, 240, \dodoi{10.1093/mnras/stw1257}

\bibitem[{{Bigiel} {et~al.}(2008){Bigiel}, {Leroy}, {Walter}, {Brinks}, {de Blok}, {Madore}, \& {Thornley}}]{bigiel2008}
{Bigiel}, F., {Leroy}, A., {Walter}, F., {et~al.} 2008, \aj, 136, 2846, \dodoi{10.1088/0004-6256/136/6/2846}

\bibitem[{{Binney} \& {Tremaine}(2008)}]{binney_tremaine}
{Binney}, J., \& {Tremaine}, S. 2008, {Galactic Dynamics: Second Edition}

\bibitem[{{Blumenthal} {et~al.}(1986){Blumenthal}, {Faber}, {Flores}, \& {Primack}}]{Blum86}
{Blumenthal}, G.~R., {Faber}, S.~M., {Flores}, R., \& {Primack}, J.~R. 1986, \apj, 301, 27, \dodoi{10.1086/163867}

\bibitem[{{Brandenburg} \& {Subramanian}(2005)}]{brandenburg2005}
{Brandenburg}, A., \& {Subramanian}, K. 2005, \physrep, 417, 1, \dodoi{10.1016/j.physrep.2005.06.005}

\bibitem[{{Butsky} {et~al.}(2017){Butsky}, {Zrake}, {Kim}, {Yang}, \& {Abel}}]{butsky2017}
{Butsky}, I., {Zrake}, J., {Kim}, J.-h., {Yang}, H.-I., \& {Abel}, T. 2017, \apj, 843, 113, \dodoi{10.3847/1538-4357/aa799f}

\bibitem[{{Chy{\.z}y} \& {Buta}(2008)}]{chyzy2008}
{Chy{\.z}y}, K.~T., \& {Buta}, R.~J. 2008, \apjl, 677, L17, \dodoi{10.1086/587958}

\bibitem[{{Crutcher} {et~al.}(2010){Crutcher}, {Wandelt}, {Heiles}, {Falgarone}, \& {Troland}}]{crutcher}
{Crutcher}, R.~M., {Wandelt}, B., {Heiles}, C., {Falgarone}, E., \& {Troland}, T.~H. 2010, \apj, 725, 466, \dodoi{10.1088/0004-637X/725/1/466}

\bibitem[{{Dacunha} {et~al.}(2025){Dacunha}, {Martin-Alvarez}, {Clark}, \& {Lopez-Rodriguez}}]{Dacunha2025}
{Dacunha}, T., {Martin-Alvarez}, S., {Clark}, S.~E., \& {Lopez-Rodriguez}, E. 2025, \apj, 980, 197, \dodoi{10.3847/1538-4357/adab72}

\bibitem[{{Daniel} \& {Wyse}(2018)}]{Daniel2018}
{Daniel}, K.~J., \& {Wyse}, R. F.~G. 2018, \mnras, 476, 1561, \dodoi{10.1093/mnras/sty199}

\bibitem[{{Elmegreen} \& {Elmegreen}(1986)}]{elmegreen1986}
{Elmegreen}, B.~G., \& {Elmegreen}, D.~M. 1986, \apj, 311, 554, \dodoi{10.1086/164795}

\bibitem[{{Evans} \& {Hawley}(1988)}]{constrained_transport}
{Evans}, C.~R., \& {Hawley}, J.~F. 1988, \apj, 332, 659, \dodoi{10.1086/166684}

\bibitem[{{Federrath} {et~al.}(2011){Federrath}, {Chabrier}, {Schober}, {Banerjee}, {Klessen}, \& {Schleicher}}]{fed_ssdynamo}
{Federrath}, C., {Chabrier}, G., {Schober}, J., {et~al.} 2011, \prl, 107, 114504, \dodoi{10.1103/PhysRevLett.107.114504}

\bibitem[{{Ferland} {et~al.}(2017){Ferland}, {Chatzikos}, {Guzm{\'a}n}, {Lykins}, {van Hoof}, {Williams}, {Abel}, {Badnell}, {Keenan}, {Porter}, \& {Stancil}}]{Cloudy}
{Ferland}, G.~J., {Chatzikos}, M., {Guzm{\'a}n}, F., {et~al.} 2017, \rmxaa, 53, 385, \dodoi{10.48550/arXiv.1705.10877}

\bibitem[{{Fletcher} {et~al.}(2011){Fletcher}, {Beck}, {Shukurov}, {Berkhuijsen}, \& {Horellou}}]{fletcher2011}
{Fletcher}, A., {Beck}, R., {Shukurov}, A., {Berkhuijsen}, E.~M., \& {Horellou}, C. 2011, \mnras, 412, 2396, \dodoi{10.1111/j.1365-2966.2010.18065.x}

\bibitem[{{Frankel} {et~al.}(2020){Frankel}, {Sanders}, {Ting}, \& {Rix}}]{Frankel2020}
{Frankel}, N., {Sanders}, J., {Ting}, Y.-S., \& {Rix}, H.-W. 2020, \apj, 896, 15, \dodoi{10.3847/1538-4357/ab910c}

\bibitem[{{Frick} {et~al.}(2016){Frick}, {Stepanov}, {Beck}, {Sokoloff}, {Shukurov}, {Ehle}, \& {Lundgren}}]{Frick2016}
{Frick}, P., {Stepanov}, R., {Beck}, R., {et~al.} 2016, \aap, 585, A21, \dodoi{10.1051/0004-6361/201526796}

\bibitem[{{Geach} {et~al.}(2023){Geach}, {Lopez-Rodriguez}, {Doherty}, {Chen}, {Ivison}, {Bendo}, {Dye}, \& {Coppin}}]{Geach2023}
{Geach}, J.~E., {Lopez-Rodriguez}, E., {Doherty}, M.~J., {et~al.} 2023, arXiv e-prints, arXiv:2309.02034, \dodoi{10.48550/arXiv.2309.02034}

\bibitem[{{Goldsmith} {et~al.}(2008){Goldsmith}, {Heyer}, {Narayanan}, {Snell}, {Li}, \& {Brunt}}]{goldsmith2008}
{Goldsmith}, P.~F., {Heyer}, M., {Narayanan}, G., {et~al.} 2008, \apj, 680, 428, \dodoi{10.1086/587166}

\bibitem[{{Gressel} {et~al.}(2013){Gressel}, {Elstner}, \& {Ziegler}}]{gressel2013}
{Gressel}, O., {Elstner}, D., \& {Ziegler}, U. 2013, \aap, 560, A93, \dodoi{10.1051/0004-6361/201322349}

\bibitem[{{Henriksen}(2017)}]{henriksen2017}
{Henriksen}, R.~N. 2017, \mnras, 469, 4806, \dodoi{10.1093/mnras/stx1169}

\bibitem[{{Hernquist}(1990)}]{Hern90}
{Hernquist}, L. 1990, \apj, 356, 359, \dodoi{10.1086/168845}

\bibitem[{{Hockney} \& {Eastwood}(1981)}]{pm}
{Hockney}, R.~W., \& {Eastwood}, J.~W. 1981, {Computer Simulation Using Particles}

\bibitem[{{Joung} \& {Mac Low}(2006)}]{Joung2006}
{Joung}, M.~K.~R., \& {Mac Low}, M.-M. 2006, \apj, 653, 1266, \dodoi{10.1086/508795}

\bibitem[{{Keller} {et~al.}(2014){Keller}, {Wadsley}, {Benincasa}, \& {Couchman}}]{keller2014}
{Keller}, B.~W., {Wadsley}, J., {Benincasa}, S.~M., \& {Couchman}, H.~M.~P. 2014, \mnras, 442, 3013, \dodoi{10.1093/mnras/stu1058}

\bibitem[{{Khoperskov} \& {Khrapov}(2018)}]{Khoperskov2018}
{Khoperskov}, S.~A., \& {Khrapov}, S.~S. 2018, \aap, 609, A104, \dodoi{10.1051/0004-6361/201629988}

\bibitem[{{Kim} {et~al.}(2006){Kim}, {Kim}, \& {Ostriker}}]{Kim2006}
{Kim}, C.-G., {Kim}, W.-T., \& {Ostriker}, E.~C. 2006, \apjl, 649, L13, \dodoi{10.1086/508160}

\bibitem[{{Kim} {et~al.}(2014){Kim}, {Abel}, {Agertz}, {Bryan}, {Ceverino}, {Christensen}, {Conroy}, {Dekel}, {Gnedin}, {Goldbaum}, {Guedes}, {Hahn}, {Hobbs}, {Hopkins}, {Hummels}, {Iannuzzi}, {Keres}, {Klypin}, {Kravtsov}, {Krumholz}, {Kuhlen}, {Leitner}, {Madau}, {Mayer}, {Moody}, {Nagamine}, {Norman}, {Onorbe}, {O'Shea}, {Pillepich}, {Primack}, {Quinn}, {Read}, {Robertson}, {Rocha}, {Rudd}, {Shen}, {Smith}, {Szalay}, {Teyssier}, {Thompson}, {Todoroki}, {Turk}, {Wadsley}, {Wise}, {Zolotov}, \& {AGORA Collaboration29}}]{Kim2014}
{Kim}, J.-h., {Abel}, T., {Agertz}, O., {et~al.} 2014, \apjs, 210, 14, \dodoi{10.1088/0067-0049/210/1/14}

\bibitem[{{Kim} {et~al.}(2020){Kim}, {Kim}, \& {Ostriker}}]{Kim2020}
{Kim}, W.-T., {Kim}, C.-G., \& {Ostriker}, E.~C. 2020, \apj, 898, 35, \dodoi{10.3847/1538-4357/ab9b87}

\bibitem[{{Korpi} {et~al.}(2010){Korpi}, {K{\"a}pyl{\"a}}, \& {V{\"a}is{\"a}l{\"a}}}]{korpi2010}
{Korpi}, M.~J., {K{\"a}pyl{\"a}}, P.~J., \& {V{\"a}is{\"a}l{\"a}}, M.~S. 2010, Astronomische Nachrichten, 331, 34, \dodoi{10.1002/asna.200911254}

\bibitem[{{Krumholz} {et~al.}(2018){Krumholz}, {Burkhart}, {Forbes}, \& {Crocker}}]{krumholz2018}
{Krumholz}, M.~R., {Burkhart}, B., {Forbes}, J.~C., \& {Crocker}, R.~M. 2018, \mnras, 477, 2716, \dodoi{10.1093/mnras/sty852}

\bibitem[{{Ledos} {et~al.}(2024){Ledos}, {Ntormousi}, {Takasao}, \& {Nagamine}}]{ledos2024}
{Ledos}, N., {Ntormousi}, E., {Takasao}, S., \& {Nagamine}, K. 2024, \aap, 691, A280, \dodoi{10.1051/0004-6361/202451139}

\bibitem[{{Liu} {et~al.}(2022){Liu}, {Kretschmer}, \& {Teyssier}}]{Liu2022}
{Liu}, Y., {Kretschmer}, M., \& {Teyssier}, R. 2022, \mnras, 513, 6028, \dodoi{10.1093/mnras/stac1266}

\bibitem[{{Lopez-Rodriguez} {et~al.}(2023){Lopez-Rodriguez}, {Borlaff}, {Beck}, {Reach}, {Mao}, {Ntormousi}, {Tassis}, {Martin-Alvarez}, {Clark}, {Dale}, \& {del Moral-Castro}}]{lopez-rodriguez}
{Lopez-Rodriguez}, E., {Borlaff}, A.~S., {Beck}, R., {et~al.} 2023, \apjl, 942, L13, \dodoi{10.3847/2041-8213/acaaa2}

\bibitem[{{Lovelace} \& {Hohlfeld}(1978)}]{lovelace1978}
{Lovelace}, R.~V.~E., \& {Hohlfeld}, R.~G. 1978, \apj, 221, 51, \dodoi{10.1086/156004}

\bibitem[{{Martin-Alvarez} {et~al.}(2022){Martin-Alvarez}, {Devriendt}, {Slyz}, {Sijacki}, {Richardson}, \& {Katz}}]{martin2022}
{Martin-Alvarez}, S., {Devriendt}, J., {Slyz}, A., {et~al.} 2022, \mnras, 513, 3326, \dodoi{10.1093/mnras/stac1099}

\bibitem[{{Martin-Alvarez} {et~al.}(2024){Martin-Alvarez}, {Lopez-Rodriguez}, {Dacunha}, {Clark}, {Borlaff}, {Beck}, {Rodr{\'\i}guez Montero}, {Jung}, {Devriendt}, {Slyz}, {Roman-Duval}, {Ntormousi}, {Tahani}, {Subramanian}, {Dale}, {Marcum}, {Tassis}, {del Moral-Castro}, {Tram}, \& {Jarvis}}]{martin2024}
{Martin-Alvarez}, S., {Lopez-Rodriguez}, E., {Dacunha}, T., {et~al.} 2024, \apj, 966, 43, \dodoi{10.3847/1538-4357/ad2e9e}

\bibitem[{{Ntormousi} {et~al.}(2020){Ntormousi}, {Tassis}, {Del Sordo}, {Fragkoudi}, \& {Pakmor}}]{ntormousi2020}
{Ntormousi}, E., {Tassis}, K., {Del Sordo}, F., {Fragkoudi}, F., \& {Pakmor}, R. 2020, \aap, 641, A165, \dodoi{10.1051/0004-6361/202037835}

\bibitem[{{Pakmor} {et~al.}(2020){Pakmor}, {van de Voort}, {Bieri}, {G{\'o}mez}, {Grand}, {Guillet}, {Marinacci}, {Pfrommer}, {Simpson}, \& {Springel}}]{pakmor}
{Pakmor}, R., {van de Voort}, F., {Bieri}, R., {et~al.} 2020, \mnras, 498, 3125, \dodoi{10.1093/mnras/staa2530}

\bibitem[{{Parker}(1955)}]{Parker1955}
{Parker}, E.~N. 1955, \apj, 122, 293, \dodoi{10.1086/146087}

\bibitem[{{Pfrommer} {et~al.}(2022){Pfrommer}, {Werhahn}, {Pakmor}, {Girichidis}, \& {Simpson}}]{pfrommer2022}
{Pfrommer}, C., {Werhahn}, M., {Pakmor}, R., {Girichidis}, P., \& {Simpson}, C.~M. 2022, \mnras, 515, 4229, \dodoi{10.1093/mnras/stac1808}

\bibitem[{{Planck Collaboration} {et~al.}(2016){Planck Collaboration}, {Ade}, {Aghanim}, {Alves}, {Arnaud}, {Arzoumanian}, {Ashdown}, {Aumont}, {Baccigalupi}, {Banday}, {Barreiro}, {Bartolo}, {Battaner}, {Benabed}, {Beno{\^\i}t}, {Benoit-L{\'e}vy}, {Bernard}, {Bersanelli}, {Bielewicz}, {Bock}, {Bonavera}, {Bond}, {Borrill}, {Bouchet}, {Boulanger}, {Bracco}, {Burigana}, {Calabrese}, {Cardoso}, {Catalano}, {Chiang}, {Christensen}, {Colombo}, {Combet}, {Couchot}, {Crill}, {Curto}, {Cuttaia}, {Danese}, {Davies}, {Davis}, {de Bernardis}, {de Rosa}, {de Zotti}, {Delabrouille}, {Dickinson}, {Diego}, {Dole}, {Donzelli}, {Dor{\'e}}, {Douspis}, {Ducout}, {Dupac}, {Efstathiou}, {Elsner}, {En{\ss}lin}, {Eriksen}, {Falceta-Gon{\c{c}}alves}, {Falgarone}, {Ferri{\`e}re}, {Finelli}, {Forni}, {Frailis}, {Fraisse}, {Franceschi}, {Frejsel}, {Galeotta}, {Galli}, {Ganga}, {Ghosh}, {Giard}, {Gjerl{\o}w}, {Gonz{\'a}lez-Nuevo}, {G{\'o}rski}, {Gregorio}, {Gruppuso}, {Gudmundsson}, {Guillet}, {Harrison}, {Helou}, {Hennebelle},
  {Henrot-Versill{\'e}}, {Hern{\'a}ndez-Monteagudo}, {Herranz}, {Hildebrandt}, {Hivon}, {Holmes}, {Hornstrup}, {Huffenberger}, {Hurier}, {Jaffe}, {Jaffe}, {Jones}, {Juvela}, {Keih{\"a}nen}, {Keskitalo}, {Kisner}, {Knoche}, {Kunz}, {Kurki-Suonio}, {Lagache}, {Lamarre}, {Lasenby}, {Lattanzi}, {Lawrence}, {Leonardi}, {Levrier}, {Liguori}, {Lilje}, {Linden-V{\o}rnle}, {L{\'o}pez-Caniego}, {Lubin}, {Mac{\'\i}as-P{\'e}rez}, {Maino}, {Mandolesi}, {Mangilli}, {Maris}, {Martin}, {Mart{\'\i}nez-Gonz{\'a}lez}, {Masi}, {Matarrese}, {Melchiorri}, {Mendes}, {Mennella}, {Migliaccio}, {Miville-Desch{\^e}nes}, {Moneti}, {Montier}, {Morgante}, {Mortlock}, {Munshi}, {Murphy}, {Naselsky}, {Nati}, {Netterfield}, {Noviello}, {Novikov}, {Novikov}, {Oppermann}, {Oxborrow}, {Pagano}, {Pajot}, {Paladini}, {Paoletti}, {Pasian}, {Perotto}, {Pettorino}, {Piacentini}, {Piat}, {Pierpaoli}, {Pietrobon}, {Plaszczynski}, {Pointecouteau}, {Polenta}, {Ponthieu}, {Pratt}, {Prunet}, {Puget}, {Rachen}, {Reinecke}, {Remazeilles}, {Renault},
  {Renzi}, {Ristorcelli}, {Rocha}, {Rossetti}, {Roudier}, {Rubi{\~n}o-Mart{\'\i}n}, {Rusholme}, {Sandri}, {Santos}, {Savelainen}, {Savini}, {Scott}, {Soler}, {Stolyarov}, {Sudiwala}, {Sutton}, {Suur-Uski}, {Sygnet}, {Tauber}, {Terenzi}, {Toffolatti}, {Tomasi}, {Tristram}, {Tucci}, {Umana}, {Valenziano}, {Valiviita}, {Van Tent}, {Vielva}, {Villa}, {Wade}, {Wandelt}, {Wehus}, {Ysard}, {Yvon}, \& {Zonca}}]{planck1}
{Planck Collaboration}, {Ade}, P.~A.~R., {Aghanim}, N., {et~al.} 2016, \aap, 586, A138, \dodoi{10.1051/0004-6361/201525896}

\bibitem[{{Ponnada} {et~al.}(2022){Ponnada}, {Panopoulou}, {Butsky}, {Hopkins}, {Loebman}, {Hummels}, {Ji}, {Wetzel}, {Faucher-Gigu{\`e}re}, \& {Hayward}}]{ponnada_2022}
{Ponnada}, S.~B., {Panopoulou}, G.~V., {Butsky}, I.~S., {et~al.} 2022, \mnras, 516, 4417, \dodoi{10.1093/mnras/stac2448}

\bibitem[{{Pudritz} \& {Silk}(1989)}]{pudritz1989}
{Pudritz}, R.~E., \& {Silk}, J. 1989, \apj, 342, 650, \dodoi{10.1086/167625}

\bibitem[{{Querejeta} {et~al.}(2024){Querejeta}, {Leroy}, {Meidt}, {Schinnerer}, {Belfiore}, {Emsellem}, {Klessen}, {Sun}, {Sormani}, {Be{\v{s}}li{\'c}}, {Cao}, {Chevance}, {Colombo}, {Dale}, {Garc{\'\i}a-Burillo}, {Glover}, {Grasha}, {Groves}, {Koch}, {Neumann}, {Pan}, {Pessa}, {Pety}, {Pinna}, {Ramambason}, {Razza}, {Romanelli}, {Rosolowsky}, {Ruiz-Garc{\'\i}a}, {S{\'a}nchez-Bl{\'a}zquez}, {Smith}, {Stuber}, {Ubeda}, {Usero}, \& {Williams}}]{Querejeta2024}
{Querejeta}, M., {Leroy}, A.~K., {Meidt}, S.~E., {et~al.} 2024, \aap, 687, A293, \dodoi{10.1051/0004-6361/202449733}

\bibitem[{{Rieder} \& {Teyssier}(2016)}]{Rieder1}
{Rieder}, M., \& {Teyssier}, R. 2016, \mnras, 457, 1722, \dodoi{10.1093/mnras/stv2985}

\bibitem[{{Rieder} \& {Teyssier}(2017)}]{Rieder2}
---. 2017, \mnras, 471, 2674, \dodoi{10.1093/mnras/stx1670}

\bibitem[{{Riquelme} \& {Spitkovsky}(2009)}]{riquelme2009}
{Riquelme}, M.~A., \& {Spitkovsky}, A. 2009, \apj, 694, 626, \dodoi{10.1088/0004-637X/694/1/626}

\bibitem[{{Robinson} \& {Wadsley}(2024)}]{Robinson2024}
{Robinson}, H., \& {Wadsley}, J. 2024, \mnras, \dodoi{10.1093/mnras/stae2132}

\bibitem[{{Ro{\v{s}}kar} {et~al.}(2008){Ro{\v{s}}kar}, {Debattista}, {Quinn}, {Stinson}, \& {Wadsley}}]{Roskar2008}
{Ro{\v{s}}kar}, R., {Debattista}, V.~P., {Quinn}, T.~R., {Stinson}, G.~S., \& {Wadsley}, J. 2008, \apjl, 684, L79, \dodoi{10.1086/592231}

\bibitem[{{Sellwood}(2024)}]{Sell24}
{Sellwood}, J.~A. 2024, \mnras, 529, 3035, \dodoi{10.1093/mnras/stae595}

\bibitem[{{Sellwood} \& {Balbus}(1999)}]{sellwood1999}
{Sellwood}, J.~A., \& {Balbus}, S.~A. 1999, \apj, 511, 660, \dodoi{10.1086/306728}

\bibitem[{{Sellwood} \& {Binney}(2002)}]{sellwood2002}
{Sellwood}, J.~A., \& {Binney}, J.~J. 2002, \mnras, 336, 785, \dodoi{10.1046/j.1365-8711.2002.05806.x}

\bibitem[{{Sellwood} \& {Masters}(2022)}]{Sellwood2022}
{Sellwood}, J.~A., \& {Masters}, K.~L. 2022, \araa, 60, \dodoi{10.1146/annurev-astro-052920-104505}

\bibitem[{{Sellwood} \& {McGaugh}(2005)}]{SM05}
{Sellwood}, J.~A., \& {McGaugh}, S.~S. 2005, \apj, 634, 70, \dodoi{10.1086/491731}

\bibitem[{{Semenov} {et~al.}(2021){Semenov}, {Kravtsov}, \& {Caprioli}}]{semenov2021}
{Semenov}, V.~A., {Kravtsov}, A.~V., \& {Caprioli}, D. 2021, \apj, 910, 126, \dodoi{10.3847/1538-4357/abe2a6}

\bibitem[{{Shukurov} {et~al.}(2019){Shukurov}, {Rodrigues}, {Bushby}, {Hollins}, \& {Rachen}}]{shukurov2019}
{Shukurov}, A., {Rodrigues}, L. F.~S., {Bushby}, P.~J., {Hollins}, J., \& {Rachen}, J.~P. 2019, \aap, 623, A113, \dodoi{10.1051/0004-6361/201834642}

\bibitem[{{Smith} {et~al.}(2017){Smith}, {Bryan}, {Glover}, {Goldbaum}, {Turk}, {Regan}, {Wise}, {Schive}, {Abel}, {Emerick}, {O'Shea}, {Anninos}, {Hummels}, \& {Khochfar}}]{grackle}
{Smith}, B.~D., {Bryan}, G.~L., {Glover}, S. C.~O., {et~al.} 2017, \mnras, 466, 2217, \dodoi{10.1093/mnras/stw3291}

\bibitem[{{Steinwandel} {et~al.}(2020){Steinwandel}, {Dolag}, {Lesch}, {Moster}, {Burkert}, \& {Prieto}}]{steinwandel2}
{Steinwandel}, U.~P., {Dolag}, K., {Lesch}, H., {et~al.} 2020, \mnras, 494, 4393, \dodoi{10.1093/mnras/staa817}

\bibitem[{{Stix}(1975)}]{stix1975}
{Stix}, M. 1975, \aap, 42, 85

\bibitem[{{Su} {et~al.}(2018){Su}, {Hayward}, {Hopkins}, {Quataert}, {Faucher-Gigu{\`e}re}, \& {Kere{\v{s}}}}]{su2018}
{Su}, K.-Y., {Hayward}, C.~C., {Hopkins}, P.~F., {et~al.} 2018, \mnras, 473, L111, \dodoi{10.1093/mnrasl/slx172}

\bibitem[{{Sun} {et~al.}(2024){Sun}, {Calzetti}, \& {Battisti}}]{Sun2024}
{Sun}, B., {Calzetti}, D., \& {Battisti}, A.~J. 2024, \apj, 973, 137, \dodoi{10.3847/1538-4357/ad6157}

\bibitem[{{Teyssier}(2002)}]{RAMSES}
{Teyssier}, R. 2002, \aap, 385, 337, \dodoi{10.1051/0004-6361:20011817}

\bibitem[{{Toomre}(1964)}]{toomre1964}
{Toomre}, A. 1964, \apj, 139, 1217, \dodoi{10.1086/147861}

\bibitem[{{Whitworth} {et~al.}(2025){Whitworth}, {Srinivasan}, {Pudritz}, {Mac Low}, {Eadie}, {Palau}, {Soler}, {Smith}, {Pattle}, {Robinson}, {Pillsworth}, {Wadsley}, {Brucy}, {Lebreuilly}, {Hennebelle}, {Girichidis}, {Gent}, {Marin}, {S{\'a}nchez Valido}, {Camacho}, {Klessen}, \& {V{\'a}zquez-Semadeni}}]{whitworth2025}
{Whitworth}, D.~J., {Srinivasan}, S., {Pudritz}, R.~E., {et~al.} 2025, \mnras, 540, 2762, \dodoi{10.1093/mnras/staf901}

\bibitem[{{Wibking} \& {Krumholz}(2023)}]{wibking2023}
{Wibking}, B.~D., \& {Krumholz}, M.~R. 2023, \mnras, 521, 5972, \dodoi{10.1093/mnras/stac2648}

\bibitem[{{Wissing} \& {Shen}(2023)}]{wissing}
{Wissing}, R., \& {Shen}, S. 2023, \aap, 673, A47, \dodoi{10.1051/0004-6361/202244753}

\bibitem[{{Young}(1980)}]{Young80}
{Young}, P. 1980, \apj, 242, 1232, \dodoi{10.1086/158553}

\bibitem[{{Zhao} {et~al.}(2024{\natexlab{a}}){Zhao}, {Pudritz}, {Pillsworth}, {Robinson}, \& {Wadsley}}]{bo2024}
{Zhao}, B., {Pudritz}, R.~E., {Pillsworth}, R., {Robinson}, H., \& {Wadsley}, J. 2024{\natexlab{a}}, \apj, 974, 240, \dodoi{10.3847/1538-4357/ad67e2}

\bibitem[{{Zhao} {et~al.}(2024{\natexlab{b}}){Zhao}, {Zhou}, {Baan}, {Hu}, {Lazarian}, {Tang}, {Esimbek}, {He}, {Li}, {Ji}, {Chang}, \& {Tursun}}]{Zhao2024}
{Zhao}, M., {Zhou}, J., {Baan}, W.~A., {et~al.} 2024{\natexlab{b}}, \apj, 967, 18, \dodoi{10.3847/1538-4357/ad3a62}

\end{thebibliography}
\bibliographystyle{aasjournal}



\end{document}